# Applications of Raman Spectroscopy in Clinical Medicine

Yaping Qi[1,2,#,*], Esther Xinyi Chen[3,#], Dan Hu[1], Ying Yang[1], Zhenping Wu[4], Ming Zheng[5], Mohammad A. Sadi[6], Yucheng Jiang[7*], Kang Zhang[3*], Zi Chen[8*], Yong P. Chen[1,2,6,9,*]

[1]Department of Engineering Science, Faculty of Innovation Engineering, Macau University of Science and Technology, Av. Wai Long, Macau SAR, China

[2]Advanced Institute for Materials Research (AIMR), Tohoku University, Sendai 980-8577, Japan

[3]Faculty of Medicine, Macau University of Science and Technology, Av. Wai Long, Macau SAR, China

[4]State Key Laboratory of Information Photonics and Optical Communications & School of Science, Beijing University of Posts and Telecommunications, Beijing 100876, China

[5]School of Materials Science and Physics, China University of Mining and Technology, Xuzhou 221116, China

[6]Department of Physics and Astronomy and Elmore Family School of Electrical and Computer Engineering and Birck Nanotechnology Center and Purdue Quantum Science and Engineering Institute, Purdue University, West Lafayette, Indiana 47907, USA

[7]Jiangsu Key Laboratory of Micro and Nano Heat Fluid Flow Technology and Energy Application, School of Physical Science and Technology, Suzhou University of Science and Technology, Suzhou, Jiangsu 215009, China

[8]Division of Thoracic Surgery, Brigham and Women's Hospital, Harvard Medical School, 75 Francis Street, Boston, MA 02115, USA

[9]Institute of Physics and Astronomy and Villum Center for Hybrid Quantum Materials and Devices, Aarhus University, Aarhus-C, 8000 Denmark

#These authors contributed equally: Yaping Qi, Esther Xinyi Chen.

*Correspondence: ypqi@must.edu.mo; jyc@usts.edu.cn; kzhang@must.edu.mo; zchen33@bwh.harvard.edu; yongchen@purdue.edu.

**Abstract**

Raman spectroscopy provides spectral information related to the specific molecular structures of substances and has been well established as a powerful tool for studying biological tissues and diagnosing diseases. This article reviews recent advances in Raman spectroscopy and its applications in diagnosing various critical diseases, including cancers, infections, and neurodegenerative diseases, and in predicting surgical outcomes. These advances are explored through discussion of state-of-the-art forms of Raman spectroscopy, such as surface-enhanced Raman spectroscopy, resonance Raman spectroscopy, and tip-enhanced Raman spectroscopy employed in biomedical sciences. We discuss biomedical applications, including various aspects and methods of ex vivo and in vivo medical diagnosis, sample collection, data processing, and achievements in realizing the correlation between Raman spectra and biochemical information in certain diseases. Finally, we present the limitations of the current study and provide perspectives for future research.

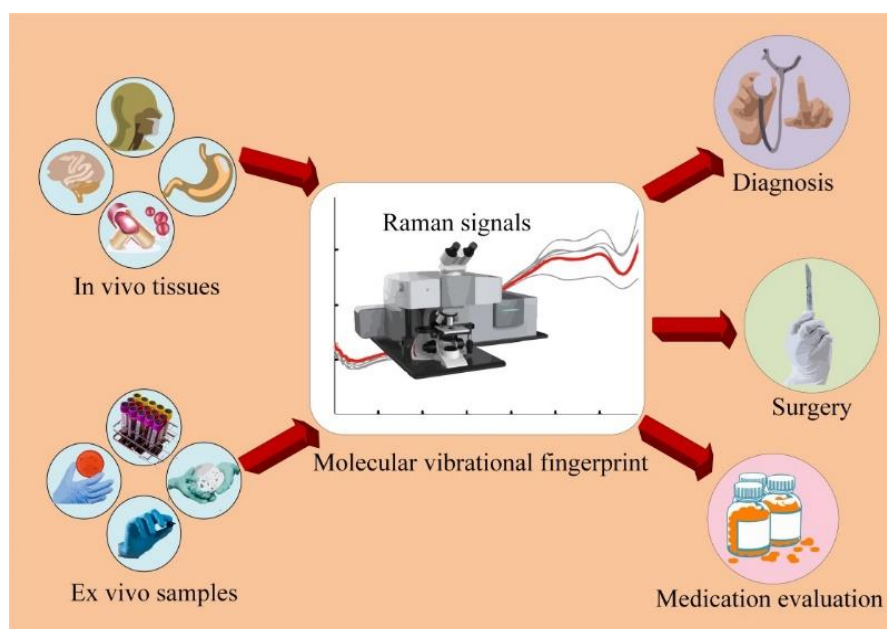

Applications of Raman spectroscopy in biomedical sciences.

# 1. Introduction

The early and accurate detection of diseases permits effective intervention. It facilitates early and effective treatment, monitoring treatment progress, slowing disease progression, and reducing mortality [1, 2]. For example, in colorectal cancer (CRC), the 5-year survival rate can reach 90% [3] with diagnosing patients at an early stage. However, only 39% of patients with CRC are diagnosed at this stage, mainly due to missed identification of lesions and the underuse of screening [4]. The same trend has also been observed in epidemiological investigations of other diseases, such as infectious [5, 6] and neurodegenerative diseases [7]. Therefore, innovative and reliable detection technologies with high specificity and sensitivity are constantly sought for disease diagnosis and severity grading [8, 9]. Optical diagnostic techniques have several advantages over other approaches, including reasonable objectivity, high speed, and low cost [10]. Among the label-free and noninvasive optical methods, Raman spectroscopy is a powerful tool that can detect the biochemical components of biological tissues. It has attracted considerable attention for clinical applications because of its excellent performance, simplicity of operation, and sensitivity in detecting and grading of lesion tissues [11-13]. Raman spectroscopy is based on Raman scattering, by which it can identify the composition of multi-component substances based on information on the characteristic molecular vibrations in the substance to be measured [14, 15]. Interestingly, all critical components of human tissues (proteins, nucleic acids, and lipids) have corresponding characteristic peaks in the Raman spectra, which contain much information. Moreover, in lesioned tissues, intracellular molecular composition and structures may vary from a normal situation and continue to change as the disease develops, which suggests that the morphology and composition of cancerous tissues can vary at the cellular and molecular levels between periods. Molecular fingerprint information and its evolution can be easily obtained with high accuracy and sensitivity using spectroscopic techniques, providing a feasible method for accurate and noninvasive detection [16]. The capability of Raman spectroscopy makes it suitable for diagnosing progressive diseases such as cancer, precancerous lesions [2-4], infectious diseases (coronavirus disease [COVID-19], dysentery, dengue fever, and epidemic hepatitis) [17-19], and metabolic diseases (osteoporosis, diabetes, etc.) [20-22]. These examples demonstrate the great potential of Raman spectroscopy in biomedical sciences. However, some limitations and challenges exist. For example, the spontaneous Raman scattering signal is weak and often subject to interference from other signals in practical applications. Obtaining a spectrum with an acceptable signal-to-noise ratio can take a long time, which is not conducive for spectral acquisition and fast imaging. Moreover, the intense fluorescence background in biological samples can further reduce the signal-to-noise ratio of Raman spectra.

Previous reviews of Raman spectroscopy in medical applications have typically focused on only one part of disease management, including diagnosis [4], guiding surgeries [13], assessing treatment results [23], or concentrated on only one or a group of certain diseases (cancers and bacterial infections, etc.), and are commonly limited to a single type of sample, such as biofluids [24], cells [25], and tissues [26]. Moreover, many previous studies on Raman applications aimed at improving instrument performance for better accuracy and sensitivity, while some of the improvement methods might not be applicable in detecting biological tissues as additional light sources that may damage samples [27]. This review systematically evaluated the applications of Raman technology in various aspects of disease management, different types of diseases, and multiple samples. We summarize the recent advances in Raman spectroscopy for ex vivo and in vivo tissue diagnoses over the past decade and explore the balance between the accuracy and feasibility of the test from a medical perspective. Finally, the current challenges, difficulties, and future perspectives of Raman spectroscopy for clinical applications are presented.

## 2. Methods

### 2.1 Use of Raman spectroscopy in ex vivo and in vivo detection

### 2.1.1 Application of Raman spectroscopy in ex vivo tissue detection

In vivo research and its clinical applications are based on ex vivo studies. In detecting ex vivo tissues with Raman spectroscopy, a typical operation is spectroscopic acquisition and analysis of ex vivo samples from humans and animals. In diagnostics, Raman spectroscopy is considered an alternative technique and complementary tool for tissue biopsy [28]. Numerous analyses using this tool have been performed on liver [22], cervical [29], brain [30, 31], lung [32], breast [33, 34], and skin [35, 36] tissues. The studies primarily aimed to identify the differences in chemical components in various tissues, and the main factor that affected the results was sample preparation. Although the non-destructive and label-free detection characteristics of Raman spectroscopy allow fresh biological samples to be used without handling procedures [13], certain preprocessing techniques are still required for retrospective studies [34, 35]. Various samples, including fresh, frozen, and formalin-fixed paraffin-embedded (FFPE) tissues, have been used for Raman spectroscopy-based pathological diagnoses [33, 34, 36]. These three sample types have different advantages and limitations. In general, a fresh tissue is an ideal sample. However, its composition and structure may change once the tissue is excised. Maintaining the sample in a state similar to that in vivo in practical experiments is difficult. A significant challenge is maintaining the state of cells to simulate living tissues. Previous studies have presented several solutions to this challenge. Malini et al. [37] soaked tissue sections in cold saline to shortly maintain the morphology of the cells and keep the sample moisturized. This study demonstrated that the Raman spectra did not display any changes from the fresh tissue spectra under these conditions. Additionally, Malini et al. considered the effect of osmolarity but did not account for the difference in pH between saline and body fluids. Fu et al. [30] conducted a similar study using phosphate buffer at a pH similar to that of body fluids instead of saline, thereby avoiding the denaturation and aggregation of biological macromolecules caused by changes in acidity or alkalinity. Subsequently, with the development of freezing technology, liquid nitrogen was used to preserve samples [32]. Meanwhile, in clinical studies, an abundant tissue bank is needed as a significant resource for accurate improvement in on-site retrospective research [35], so reliable preprocessing is aimed at the long-term preservation of samples. However, the current FFPE operation may affect the biochemical composition and lead to changes in Raman spectra. Various chemical and digital dewaxing methods can remove paraffin peaks from the tissue Raman spectra [33, 38, 39, 40]. Because the tissue preparation and dewaxing process also eliminate some diagnostic biomarkers, these operations may add difficulties to the classification of tissues for clinical identification and thereby the identification of FFPE samples typically has lower precision than that of frozen tissues. Recently, Ning et al. [40] elaborated on the loss of biochemical tissue information during sample preparation. The experimental results suggest that although the chemical treatment process misled the tissue spectral analysis, some commonly used multivariate analysis methods, including principal component analysis, linear discriminant analysis (PCA-LDA), and partial least squares discriminant analysis (PLS-DA), effectively distinguished between healthy tissue, ductal carcinoma in situ, and invasive ductal carcinoma. Therefore, although the fresh or frozen tissue preserved comprehensive qualitative and quantitative compositional information, the spectral results obtained from dewaxed tissues were sufficient to distinguish between normal and lesioned tissues.

Furthermore, tissue smears are becoming popular in clinical diagnosis, as more biomarkers are identified in biofluids that are linked to specific diseases. Recently, biofluid smears have been used for Raman-based diagnostics, including the blood [41, 42], saliva [43], respiratory secretion [44], cervical tissue [45], and tear smears [46, 47]. Generally, smear tests offer new noninvasive or minimally invasive detection options as they allow detection with sparse biological tissue and require less tissue than slices. However, the development of smear substrates is a complicated and lengthy process, and researchers are constantly attempting to develop substrates with low spectral impact [48-49]. Additionally, substrates that provide consistent Raman spectra, irrespective of the excitation wavelength, are costly and beyond the reach of many laboratories in less-resourced settings. To reduce the effect of the substrate on the blood smear spectrum, Otange et al. [48] smeared a conductive silver paste consisting of a mixture of solvent resins and silver metal particles on glass slides (with Ag being 35%–65% per weight), achieving excellent accuracy. Similarly, Birech et al. [49] have reported low-cost smear substrates with conductive silver paint for Raman spectroscopic screening of metabolic diseases in whole blood. Raman signals from blood applied onto a conductive silver paint-smeared glass slide were enhanced by a factor of 1.7 compared to those from a thick blood smear on the glass. Their results suggested that paste smear background signals emanating from these substrates had little influence on the Raman signals of blood samples and suppressed the photoluminescence signals from glass. They also used this smear for Raman spectroscopy to screen blood samples for type 2 diabetes [50]. Moreover, other excellent materials that achieve low interference and repeatability are available, such as Raman-grade calcium fluoride [51], aluminum-coated substrates [52], magnesium fluoride, potassium bromide, and sodium chloride. However, elemental fluoride and fluoride ions are also highly toxic as they disrupt cell enzymatic processes, such as the transformation of carbohydrates and lipids and synthesis of hormones, thus inhibiting tissue respiration.

### 2.1.2 Application of Raman spectroscopy for in vivo tissue detection

Raman spectroscopy is not only highly accurate for detecting biological tissues but also provides robust data and a technical foundation for in vivo biological studies. Biological in vivo testing uses a living individual to analyze the physical or chemical interactions between the testing instrument and the whole living organism or the local area [11]. This method can obtain realistic results without affecting the physiological state of an organism.

The application of Raman spectroscopy to biological in vivo testing has been developed based on ex vivo experiments. Raman spectroscopy can be applied directly to live tissue detection without processing or marker injection [11]. In this process, in vivo Raman data collection is achieved using a portable Raman system and Raman probes. These data can be used to construct diagnostic models and validate the application of Raman-based in vivo detection. The current application of Raman spectroscopy in clinical in vivo detection involves two main approaches. The first is a minimally invasive approach using a Raman system combined with medical endoscopy to achieve in situ detection [53-59]. This method detects and measures body tissues, including the respiratory system [54, 55] and digestive system [56-59]. The second is the direct detection of living tissue using the Raman system, as the endoscope may not reach the lesion sites [60-64].

## 2.2 Raman spectra acquisition enhancement techniques

Raman spectroscopy-assisted endoscopy has gradually become a complete system, based on previous studies, especially image-guided Raman endoscopy. Consequently, different types of Raman signal enhancement techniques have been developed to improve the intensity of Raman signals [65]. For example, surface-enhanced Raman spectroscopy (SERS) [66-69], resonance Raman spectroscopy (RRS) [70], and tip-enhanced Raman spectroscopy (TERS) [71-74] are often used to enhance the sensitivity of detection and/or improve the spatial resolution. TERS, which can detect the vibration of a single molecule, has been widely employed for the optical analysis of biological tissue [71-74]. These novel techniques demonstrate high biochemical sensitivity and selectivity, indicating that they are promising for biological and medical applications. Table 1 summarizes the principles, advantages, and limitations of five commonly used Raman techniques.

Table 1: Principles, advantages, and limitations of the five basic Raman techniques used to give enhanced Raman signals, under two classifications.

| Classifications | Types | Theoretical foundations | Advantages | Disadvantages | References |
|---|---|---|---|---|---|
| Coherent Raman Spectroscopy | CARS | Four-wave mixing | Almost no fluorescence interference; high imaging sensitivity and speed; strong signal intensity | Signal is affected by a co-generated coherent background signal | [75-77] |
| | SRS | Four-wave mixing | Superior in maintaining undistorted Raman spectra; high imaging sensitivity and speed; strong signal intensity; low detection limit | Signal is affected by co-generated coherent background signals | [78-80] |
| | RRS | Resonance effect | Suitable for biological chromophores; high signal-to-noise ratio; selective signal enhancement | Chemical groups that do not participate in the electronic transition cannot be observed; Fluorescence interference | [80, 81] |
| Enhanced Raman Spectroscopy | TERS | Local optical, electromagnetic field enhancement | Allow detection in tiny feature sizes; high sensitivity | Diffraction-limited spatial resolution; low stability | [82, 83] |
| | SERS | Chemical enhancement via charge Transfer; Electromagnetic enhancement | High probability of obtaining Raman enhancement; high sensitivity; low detection limit | Stringent requirements for analytes and substrates | [82, 84, 85] |

(CARS: Coherent Anti-Stokes Raman Scattering, SRS: Stimulated Raman Spectroscopy, RRS: Resonance Raman Spectroscopy, TERS: Tip-Enhanced Raman Spectroscopy, SERS: Surface-Enhanced Raman Spectroscopy)

Previously, flexible endoscopes [86] based on white light reflectance were the standard instruments for the detection of cancer and surveillance of precancers in internal organs [87]. However, one clinical challenge is the detection of premalignant lesions and early neoplastic changes. Reliance on subjective visual diagnostic criteria (i.e., structural, and morphological tissue details) results in poor diagnostic accuracy because of the lack of noticeable morphological changes associated with early neoplastic transformation, even for experienced endoscopists [88]. New techniques to improve detection accuracy and sensitivity by observing endogenous fluorophores in tissues and enhancing the image contrast of the tissue microvasculature are beginning to be applied clinically. One outstanding result is narrow-band imaging (NBI) [78-80]. Huang et al. [89] used this NBI-guided Raman spectroscopy to diagnose gastric dysplasia in vivo. A diagnostic sensitivity of 94.4% and specificity of 96.3% were achieved using the Raman spectral differences between normal and dysplastic gastric tissues. Moreover, they identified that albumin, nucleic acids, phospholipids, and histones were the most significant features in constructing the diagnostic model [90]. Bergholt et al. [91] have reported the implementation of NBI-guided Raman endoscopy for the first time. The results indicated that significant Raman spectral differences reflecting the distinct composition and morphology in the nasopharynx and larynx should be essential parameters in the interpretation. However, the specificity and sensitivity of these two studies differed significantly, which was most likely caused by the intensity difference of the Raman signals in different human tissues.

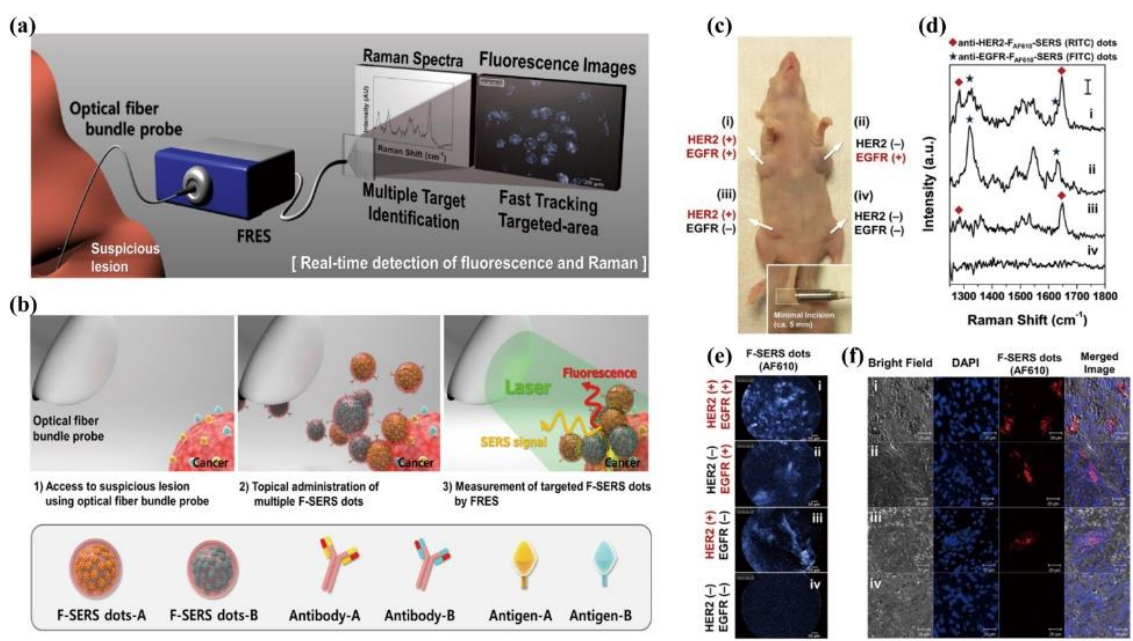

Figure 1. (a–b): Schematic illustration of real-time multiplexed imaging using the fluorescence-Raman endoscopic system (FRES) and surface-enhanced Raman scattering nanoprobes (F-SERS dots). (a) The real-time fluorescence imaging tracks the locations of the probe-targeted areas, and a concurrent SERS spectral analysis identifies the species of targets. (b) Illustration of the in vivo multiplexed molecular imaging procedure. (c–f): Demonstration of an in vivo active targeting ability of the F-SERS dots on the human epidermal growth factor receptor 2 (HER2) and epidermal growth factor receptor (EGFR) positive breast tumor xenografts. (c) A photograph of the tumor-bearing mouse with the receptor expression status. The real-time fluorescence images and Raman spectra were simultaneously obtained with an optical fiber bundle probe of the FRES (lower box). (d) The Raman spectra were obtained by the FRES at a laser power of 2.7 mW and an acquisition time of 1 s. The observed Raman bands in the Raman spectra correspond to the [(rhodamine B isothiocyanate)] RITC (◆) and [fluorescein isothiocyanate (FITC)] (★) from the F-SERS dots. (e) Fluorescence images were obtained by the FRES in real-time (12 frames/s). The bright area in fluorescence images corresponds to the targeted areas. (f) The confocal fluorescence laser scanning images of the tumor sites. The nuclei of the tumor cells stained with 4´,6-diamidino-2-phenylindole (DAPI) dye are presented as blue spots,

and the targeted F-SERS dots containing [Alexa Fluor (AF) 610] are indicated by red spots. Reproduced from Ref. [53] with permission from Nature Portfolio, copyright 2015.

Autofluorescence imaging (AFI) is another promising, wide-field imaging modality. However, these wide-field imaging modalities still suffer from insufficient diagnostic specificity owing to a lack of ability to reveal specific biomolecular information regarding the tissue. The integration of Raman-based technologies with other optical modalities provides an excellent solution to overcome this limitation. For example, Lin et al. [92] developed an integrated four-modality endoscopy system combining white-light interferometry, AFI, diffuse reflectance spectroscopy, and Raman spectroscopy technologies for in vivo endoscopic cancer detection, which achieved both high diagnostic sensitivity (98.6%) and high specificity (95.1%) for differentiating cancer from normal tissue sites. Jeong et al. [53] developed a dual-modal fluorescence-Raman endomicroscopic system that combines fluorescence and SERS nanoprobes (Figure 1a, b). This system was utilized to simultaneously detect two biomarkers, human epidermal growth factor receptor two and epidermal growth factor receptor, in a breast cancer orthotopic model (Figure 1c–f). Kim et al. [93] further demonstrated the capability of a fluorescence-Raman endomicroscopic system for diagnosing CRC in an orthotopic xenograft model (Figure 2). Developing minimally invasive or noninvasive advanced optical technologies based on the intrinsic biomolecular signatures of cells and tissues would represent a cornerstone in endoscopic diagnostics.

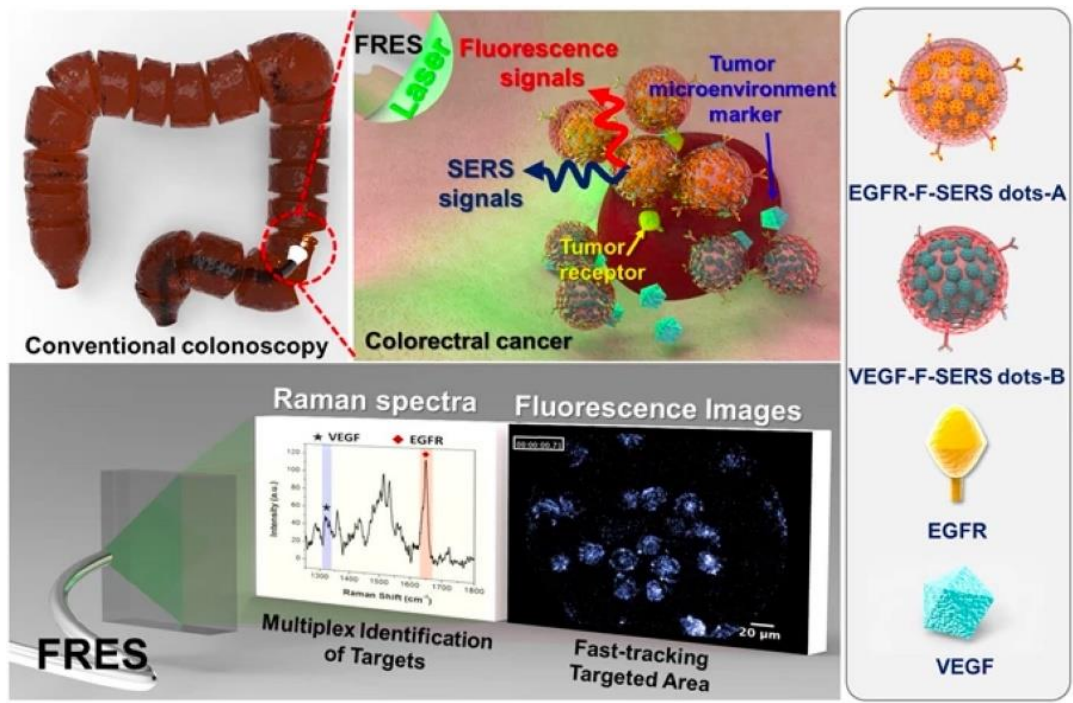

Figure 2. Schematic illustration of the in vivo multiplex molecular diagnosis on colorectal cancer using simultaneous fluorescence-Raman endoscopic system (FRES). FRES detected fluorescence and Raman signals simultaneously for the molecular characterization of a tumor. When antibody-conjugated F-SERS dots were sprayed onto HT29-efflux colon cancer cells, the antibody-conjugated F-SERS dots bound to the colon cancer cells EGFR and tumor microenvironments VEGF. FRES simultaneously utilizes the fluorescence signal of AF 610 for fast signal detection, and the Raman signals for multiplex targeting from the silver nanoparticles labeled by two kinds of Raman active compounds [(RITC, -A) and (FITC, -B)]. Reproduced from Ref. [93] with permission from Nature Portfolio, copyright 2017.

Generally, a contact device requires fewer detection sites than an endoscope does. The high flexibility of the contact Raman system allows the accurate prediction of the contour of the lesion site. It presents excellent prospects for evaluating treatment results and optimizing surgical protocols. For example, Desroches et al. [94] conducted a detailed characterization of a handheld Raman spectroscopy system to maximize the volume of resected cancer tissue during glioma surgery. Preliminary measurements of normal, necrotic, and cancerous tissues collected from ten patients demonstrated that necrosis could be distinguished from vital tissues, including regular and cancerous brain tissues, with an accuracy of 87%. Similarly, Jermyn et al. proposed a hand-held contact Raman spectroscopy probe technique for live, local detection of cancer cells in the human brain [63]. Using this technique, Jermyn et al. precisely identified cancer cells, with 93% sensitivity and 91% specificity. Based on these impressive results with excellent accuracy, a surgical plan can be developed based on the contour and area of the lesion before open surgery. Additionally, the results can be evaluated by examining the margins of the tissue in situ after lesion removal.

More prospective studies have successfully established in vivo using animal models. In most situations, simultaneous detection of multiple biomarkers at an early stage offers additional advantages, such as improved diagnostic accuracy and treatment response. Compared with the commonly used fluorescence methods, the combination of handheld Raman spectroscopy and surface-enhanced resonance Raman scattering nanoparticles (SERRS NPs) overcomes the limitations of spectral overlapping and strong background autofluorescence [95, 96]. For instance, Huang et al. [97] evaluated the ability of a handheld Raman scanner guided by SERS nanoparticles to identify the extent of microscopic tumors in a genetically engineered RCAS/tv-a glioblastoma mouse model. Although this study demonstrates the possibility of combining gold-silicon dioxide SERS nanoparticles with a handheld Raman scanner to guide surgical resection, the entire SERS image could still not be acquired. However, in vivo multiplex detection of actively targeted biomarkers, which is challenging, was achieved. Using SERS nanoparticles capable of three multiplexing, Dinish et al. [98] demonstrated that they can actively target the in vitro and in vivo detection of three intrinsic cancer biomarkers, including EGFR, CD44, and TGFβ II, in a breast cancer model. Compared to Karabeber et al. [99], Dinish et al. [98] injected SERS NP into the tumor center rather than the veins to avoid the first-pass clearance effect of the liver to attenuate the SERS signal.

## 2.3 Data analysis and machine learning

Researchers have become increasingly interested in machine learning-assisted Raman spectroscopy analysis for biomedical applications such as diagnosis, surgery, and disease treatment. Table 2 summarizes the recent advances in combining machine learning methods with Raman spectroscopy for these applications. Machine learning combined with Raman spectroscopy for medical applications is an emerging research area in the health sciences. The accuracy of the information provided by Raman spectral data depends on the analysis and processing of the spectral data. Therefore, some basic preprocessing methods for spectral data, such as removing the fluorescence background, reducing noise, and correcting the baseline, remain important [100]. Additionally, spectra normalization (e.g., spectral area, maximum value, average intensity normalization) can reduce the effects of fluctuations in excitation intensity and better compare spectral shapes and relative peak intensities between different tissues. The analysis of the Raman spectra mainly includes two steps: spectral feature extraction and tissue identification. The diagnosis result may be affected if we analyze the entire spectral data set because the initially collected data may contain considerable invalid and interfering information.

Table 2. Recent Advances in combining machine learning methods with Raman spectroscopy for biomedical applications: diagnosis, surgery, and disease treatment.

| Applications | ML Algorithms | References |
| --- | --- | --- |
| Screening of cerebral ischemia and cerebral infarction | PCA, PLS, MRMR, SVM, KNN, PNN, DT | Fan et al. [101] (2022) |
| Classify the types of Isocitrate dehydrogenase mutations in gliomas | XGBoost, RBF-SVM | Sciortino et al. [102] (2021) |
| Classification of glioma biopsies | RF, GB | Riva et al. [103] (2021) |
| Alzheimer's disease (AD) diagnosis based on saliva analysis | ANN | Ralbovsky et al. [104] (2019) |
| Rapid screening of AD | SVM, RF, XGBoost, CatBoost | Wang et al. [105] (2022) |
| Effective primary screening of COVID-19 by serum Raman spectroscopy | SVM | Yin et al. [106] (2021) |
| Detection of COVID-19 infection by Raman spectroscopy of saliva | MILES | Ember et al. [107] (2022) |
| Classify breast cancer subtypes | PCA-DFA, PCA-SVM | Zhang et al. [108] (2022) |
| Breast cancer diagnosis | Ant colony optimization, QDA | Fallahzadeh et al. [109] (2018) |
| Classify normal and cancerous breast tissue | CNN | Ma et al. [110] (2021) |
| Diagnosis of lung cancer | CNN | Qi et al. [111] (2021) |
| Lung cancer diagnosis based on the Raman spectra of exosome | ResNet based deep learning model | Shin et al. [112] (2020) |
| Screening of ovarian cancer | BPNN, PCA | Chen et al. [113] (2022) |
| Predict gastric cancer | CNN, RF, SVM, KNN | Li et al. [114] |

| Identification of kidney tumor tissue | SVM | (2021) He et al. [115] (2021) |

(PCA: Principal Component Analysis, PLS: Partial Least Squares, MRMR: Minimum Redundancy Maximum Relevance, SVM: Support Vector Machine, KNN: K-nearest Neighbor, PNN: Probabilistic Neural Network, DT: Decision Tree, XGBoost, eXtreme Gradient Boosting, RBF: Radial Basis Function, RF: Random Forest, GB: Gradient Boosting, ANN: Artificial Neural Network, CatBoost: Categorical Boosting, MILES: Multiple Instance Learning via Embedded Instance Selection, DFA: Discriminant Function Analysis, QDA: Quadratic Discriminant Analysis, CNN: Convolutional Neural Network, BPNN: Backpropagation Neural Network).

Each Raman peak represents the corresponding Raman shift and intensity. The substance components must be identified by attributing and comparing the spectral feature peaks that represent the corresponding Raman shifts and intensities. Machine learning methods are suitable for capturing complex information from spectral data, and machine learning models can be used to identify the features of Raman spectra and classify substances [116]. Sciortino et al. [102] examined mutations in isocitrate dehydrogenase (IDH) in gliomas. They extracted 2073 Raman spectra from 38 tumor tissues and screened 103 Raman shifts using an analysis of variance. The authors employed a support vector machine (SVM) with a radial basis function kernel (RBF-SVM) and eXtreme Gradient Boosting as classification models and used the intensity of each of these 103 Raman shifts as input features [102]. Using cross-validation loops, it was determined that 52 of these shifts had the best discriminatory ability to distinguish between IDH-mutated and IDH wild-type mutations. The experimental results indicated that the RBF-SVM achieved a correct classification accuracy of 87% and XGB of 85%. Similarly, Riva et al. [103] identified 135 Raman shifts as feature inputs and used gradient boosting and random forest models to classify glioma biopsies.

However, preprocessing and feature extraction are time-consuming and tedious when dealing with large-scale data. Deep learning is an excellent option to solve this problem [116]. Deep learning is a branch of machine learning and an end-to-end neural network that combines feature extraction and classification. Because all the layers of neural networks are trained together, deep learning models can automatically extract features and yield results [116]. In recent studies, convolutional neural networks (CNN) have become popular deep learning models for Raman spectral analysis. Liu et al. [117] evaluated several classical CNN architectures applied to Raman spectral data, including LeNets (ResNet). Ma et al. [110] classified healthy and cancerous breast tissue using Raman spectroscopy combined with a CNN and achieved an overall accuracy of 92%. This experiment demonstrated that the CNN algorithm is more accurate than conventional machine learning algorithms when using a large dataset. Qi et al. [111] transformed 1D Raman data from lung tissues into 2D Raman spectra using a short-time Fourier transform and then used a CNN to classify and diagnose lung cancer. This study compared the classification accuracy of the CNN model with other machine learning models, such as PCA-LDA and SVM, where CNN, PCA-LDA, and SVM achieved an accuracy of 96.5%, 90.4%, and 93.9% in the test group, respectively [111]. This study also proved that the deep learning model can significantly improve the performance of large data samples. The application of deep learning algorithms provides innovative ideas for Raman exploration of classification tasks, making this interdisciplinary area a current research topic.

## 3. Applications of Raman spectroscopy in medical sciences

### 3.1 Cancer and precancerous lesions

Cancer remains a leading cause of death and a critical barrier to increasing life expectancy worldwide [118]. Efficacious cancer treatments rely on early detection and accurate diagnosis. Various Raman spectroscopy studies on cancer have reported on the entire process of cancer diagnosis and treatment, including differentiation of precancerous tissue, cancerous tissue, and normal tissue in ex vivo samples [119], tumor staging [120], and predictions of surgical margins [121]. Here, we review its use in breast cancer, lung cancer, and CRC.

### 3.1.1 Breast cancer

With 2.26 million new breast cancer cases in 2020, breast cancer has officially replaced lung cancer as the most prevalent cancer worldwide [118]. Numerous studies have investigated the application of Raman spectroscopy for chemical changes in cancerous tissues [119]. In a few early studies, the Raman spectra of breast tissues were fitted to those of individual breast tissue components, including fat, collagen, cell nucleus, epithelial cell cytoplasm, calcium oxalate, calcium hydroxyapatite, cholesterol-like lipid deposits, and β-carotene. Based on this theory, a later study by Abramczyk et al. [123] revealed that the main differences between normal and cancer tissues were in the spectral regions associated with the vibrations of carotenoids, fatty acids, and proteins. Lyng et al. [124] discriminated between breast cancer and benign tumors using Raman spectroscopy, and PCA was conducted to determine whether the spectra could be differentiated to their overall class of benign and cancerous tumors. The study indicated that several vibration modes significantly differed between benign and cancerous types, indicating the potential quantification of Raman spectroscopy for carcinoma grading.

Applying machine learning algorithms reduces the time required to extract and analyze the collected Raman spectra. The application of machine learning to Raman spectral classification of breast cancer has led to positive outcomes. Fallahzadeh et al. [109] combined Raman spectra with an ant colony optimization algorithm for breast cancer diagnosis and achieved a classification accuracy of 87.7% for normal, benign, and cancerous groups by selecting five features. Furthermore, Ma et al. [110] combined Raman spectroscopy with CNN to simplify the analysis process, applied a self-built 1D-CNN model to classify the spectral data of healthy and cancerous breast tissues, and compared the specificity and sensitivity of various spectral classifiers (1D-CNN, SVM, and FDA). The results reveal the potential advantages of the classification performance of the 1D-CNN model over other classification algorithms. Moreover, machine-learning-assisted Raman spectroscopy has also been used for breast cancer prevention. Yala et al. [125] developed a deep learning model based on full-field mammograms and traditional risk factors, which was more accurate than the Tyrer–Cusick model (version 8), which is the current clinical standard. The precise risk assessment provided by this pioneering research may benefit patients when traditional risk factors such as family history are unavailable.

As Raman technology allows for minimally invasive and noninvasive testing, it facilitates the examination of body fluids or metabolites as a novel and effective method to diagnose breast cancer. For example, Pichardo-Molina et al. [126] investigated serum samples from patients with breast cancer and demonstrated that Raman spectroscopy is useful for minimally invasive diagnostics. Although these studies obtained specific bands that indicated group differences and could be used as potential screening markers for breast cancer, the underlying molecular mechanism of these differences between tissues at different stages remains unknown. Similarly, a series of later studies achieved metabolite profiling of human blood using Raman spectroscopy for surgery assessment or tumor screening in breast cancer. Lin et al. [127] profiled blood samples from breast cancer patients at different treatment stages (pre- and post-surgery) based on the relative concentrations of metabolites of patients postoperatively and preoperatively. Label-free SERS technology has been used to evaluate the effects of surgery on breast cancer. Nargis et al. [128] analyzed Raman spectroscopy from the same set of serum samples from breast cancer patients using SERS to distinguish different stages of breast cancer.

### 3.1.2 Lung cancer

Lung cancer is the leading cause of cancer-related deaths (18.0%) [118]. Despite advances in surgical, radiotherapeutic, and chemotherapeutic treatments, the long-term survival rate remains low (5% at 10 years for non-small cell lung cancer) [129]. To date, the most recognized reduction in lung cancer mortality rates is related to early-stage diagnosis, followed by surgical resection. However, a high rate of false-positive results in screening using computed tomography is a significant challenge in the current detection method. Hence, better risk stratification using prediction models or biomarkers of lung cancer risk, as well as a better understanding of the biological characteristics of aggressive cancers, are required to maximize the benefit of screening. With Raman technology becoming a popular and efficient diagnostic tool, recent studies have confirmed that Raman data from many samples provide valuable information for diagnosis [130]. Here, we discuss Raman data from saliva, urine, blood, and exosome samples, thus presenting four types of models.

Human saliva contains abundant proteins and metabolites that allow for the diagnosis of certain diseases. Interestingly, Li et al. [131] measured and differentiated saliva SERS readings from 21 patients with lung cancer and 20 healthy volunteers. The results indicated that most of the Raman peak intensities decreased in patients with lung cancer compared to those in the general population. These peaks were assigned to proteins and nucleic acids, which indicated a corresponding decrease in these substances in the saliva. A recent study by Qian et al. [132] analyzed saliva samples from 61 patients with lung cancer and 66 healthy controls using a SERS system and a nano-modified chip. It summarizes 12 characteristic peaks of the spectral line in patients with lung cancer with an explanation of biochemical changes. Ke et al. [133] pooled and analyzed data obtained from relevant diagnostic studies in 2020. The pooled sensitivity and specificity in the saliva samples were 0.91 (95% CI 0.80–0.96) and 0.95 (95% CI 0.73–0.99), respectively, which indicated that Raman spectroscopy data in saliva samples provide accurate, sensitive guidance for early-stage lung cancer diagnosis. However, the susceptibility of saliva samples to sputum interference poses a challenge as it may bias the results.

Urine contains approximately 3,000 proteins and is an ideal sample for protein and peptide biomarker studies. Several researchers have reported that urine proteome profiling can predict lung cancer in control cases and other tumors [134]. Further studies on metabolomic analyses have confirmed that metabolites in urine can be used to diagnose lung cancer and assess prognosis [135]. Carrola et al. [136] analyzed and verified several main metabolites contributing to lung cancer discrimination, including 18ippurate trigonelline (reduced in patients), β-hydroxyisovalerate, α-hydroxyisobutyrate, N-acetylglutamine, and creatinine (elevated in patients relative to controls). Yang et al. [137] achieved the facile and label-free detection of lung cancer biomarkers (adenosine) in urine using magnetically assisted SERS. Furthermore, the proposed SERS sequence allows high-throughput detection with high sensitivity. Despite this research suffering from the limitation of the metabolite category, the results demonstrate the valuable potential of metabolomics for identifying putative biomarkers of lung cancer in urine.

Previous studies have investigated plasma and serum levels of deoxyribonucleic acid (DNA) and ribonucleic acid (RNA) to detect the presence of cancer [138]. Because the Raman spectra of biochemical compounds produce high-dimensional multivariate datasets, many studies have confirmed the ability of different blood component samples to detect lung cancer. Guo et al. [139] developed a highly selective detection system combining asymmetric polymerase chain reaction (PCR) and SERS for the evaluation of EGFR mutation genes in circulating tumor DNA (ctDNA) in blood samples. In addition to venous whole blood samples, serum is the most used sample, as it is a cleaner sample typically free of cells and platelets [140, 141-143]. Another study by Wang et al. [143] measured the Raman spectra of the sera of peripheral venous blood with a Micro Raman spectrometer for individuals from five groups, including healthy volunteers and patients with non-small cell lung cancer in different stages. The study suggested that the Raman spectral intensity was sequentially reduced in serum samples from the control, stage I, stage II, and stage III/IV groups.

Compared to the abovementioned body fluids, exosomes present in these biofluids are considered more efficient biomarkers [144]. Owing to their natural location, membrane proteins of exosomes can be detected without exosome lysis [145]. Increasing evidence suggests that SERS can be used for exosome detection. Such detection can be divided into label-free exosome detection [146-148] and exosome detection with SERS tags [149-150].

In general, label-free assays minimize the impact of target molecules on cells or tissues and allow information to be revealed in the native state of the sample. Additionally, although disease-associated EV-specific markers are unavailable, label-free SERS can identify unique patterns of disease-associated extracellular vesicles (EVs) through their molecular fingerprints [151]. Park et al. [146]

have reported a label-free classification method for exosomes to detect lung cancer with 95.3% sensitivity and 97.3% specificity by combining SERS and statistical pattern analysis (Figure 3). They explained this difference by listing 11 points in SERS signals from lung cancer cell-derived exosomes. Even though the classification is not precise for an accurate blood sample containing many exosomes of different origins, and only a few exosomes are derived from cancer cells, trends have been established. Therefore, this serves as a motivation for the subsequent research discussed below.

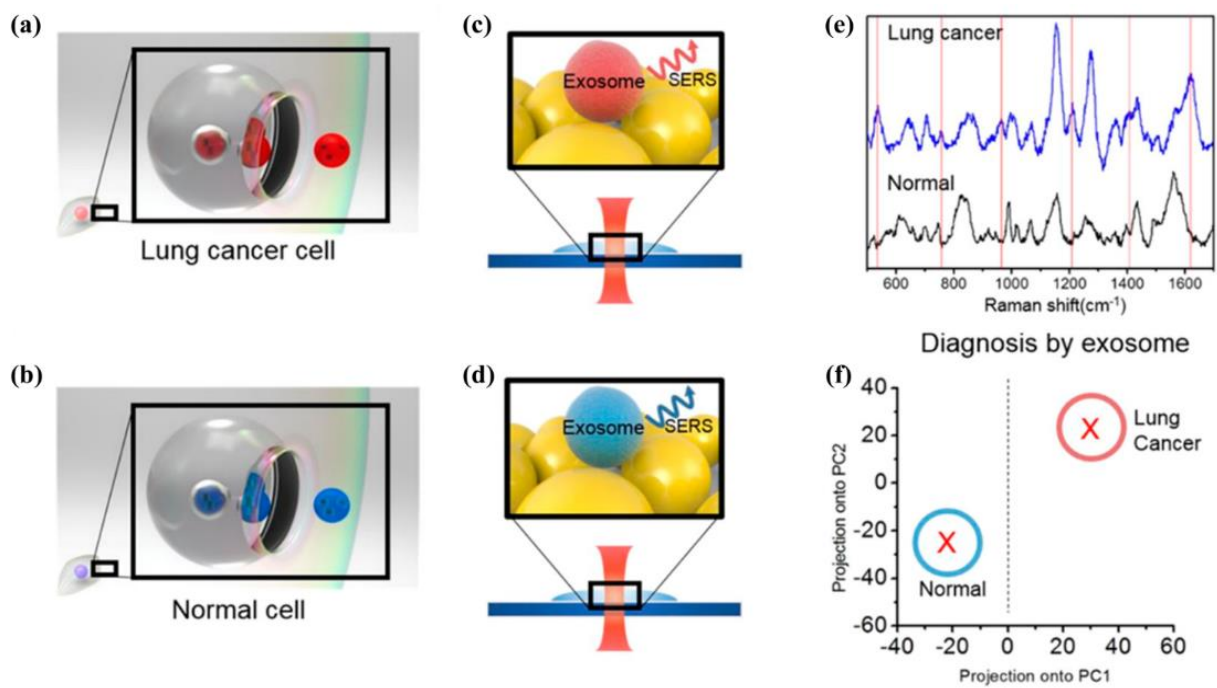

Figure 3. Schematic diagram of lung cancer diagnosis by surface-enhanced Raman spectroscopy (SERS) classification of the exosome. (a, b) Lung cancer cells and normal cell release exosomes to the extracellular environment having their own profiles by fusing multivesicular endosomes to plasma membrane, respectively. (c, d) Raman spectra of lung cancer cell and normal cell derived exosomes were achieved by SERS respectively. (e) SERS spectra achieved by methods of panels c and d are shown. The red lines indicate specific peaks of lung cancer derived exosomes. (f) Exosome classification is done by principal component analysis of SERS spectra. Identifying the origin of the exosome is done by checking the location of the plotted ×-marked dots. Dot plotted inside of the red circle indicates lung cancer cell derived exosome, and dot plotted inside of the blue circle indicates normal cell derived exosome. Reproduced from Ref. [146] with permission from ACS Publications, copyright 2017.

SERS tags, typically known as SERS-active nanoprobes, produce strong characteristic Raman signals and can be used to indirectly sense target molecules using laser Raman spectrometry or SERS microscopy, which demonstrate extraordinary features for bioanalysis. Several exosome analyses utilizing SERS have been recently reported. Shin et al. [149] analyzed specific surface protein compositions of exosomes to diagnose cancer. This study suggested that the Raman bands of exosomes from non-small cell lung cancer cells correlated well with several protein markers, including CD9, CD81, epithelial cell adhesion molecule, and EGFR.

### 3.1.3 CRC

CRC induced an estimated 1.9 million incidence cases and 0.9 million deaths worldwide in 2020 as the third most common malignancy and the second most deadly cancer [122]. Current diagnostic tests rely on histopathological analysis of tissue biopsies and suffer from limitations in their moderate diagnostic performance, invasiveness, and costly and laborious methodologies. Although Raman spectroscopy is not yet used in routine clinical diagnosis of CRC, evidence of its strengths and prospects is increasing.

As mentioned in the previous section, Raman-based endoscopes are often used in prospective studies to detect digestive tract tissues [56-59]. Generally, an in vivo probe comprises a light source and optical fiber probe. In this case, the probe can act as both a light source and detector relaying a signal. The probes used in early studies were not practical for clinical diagnosis because of their long spectral acquisition times. Moreover, some of the materials that the probes are manufactured from have large Raman cross-sections, resulting in design signal-to-noise ratio issues. However, with the development of probes over several decades, some recent studies have made progress in designing probes for gastrointestinal tissues using Raman-based endoscopy tests [152-153]. In particular, Jayhooni et al. [153] developed an endoscopic Raman spectroscopy device with side-viewing functionality that enables circumferential scanning spectral measurements inside thin conduits (Figure 4). The proposed device performed well in the detection of chemicals, harvested animal lung tissue ex vivo, a murine colon model in situ, and human skin in vivo. All the results demonstrated excellent agreement with the reported reference data, while revealing >99% wavenumber accuracies. The advent of the side-view probe has the potential to address the limitations of conventional Raman endoscopy systems in terms of angle and detection range, as well as contributing to the screening of lesions in narrow lumens.

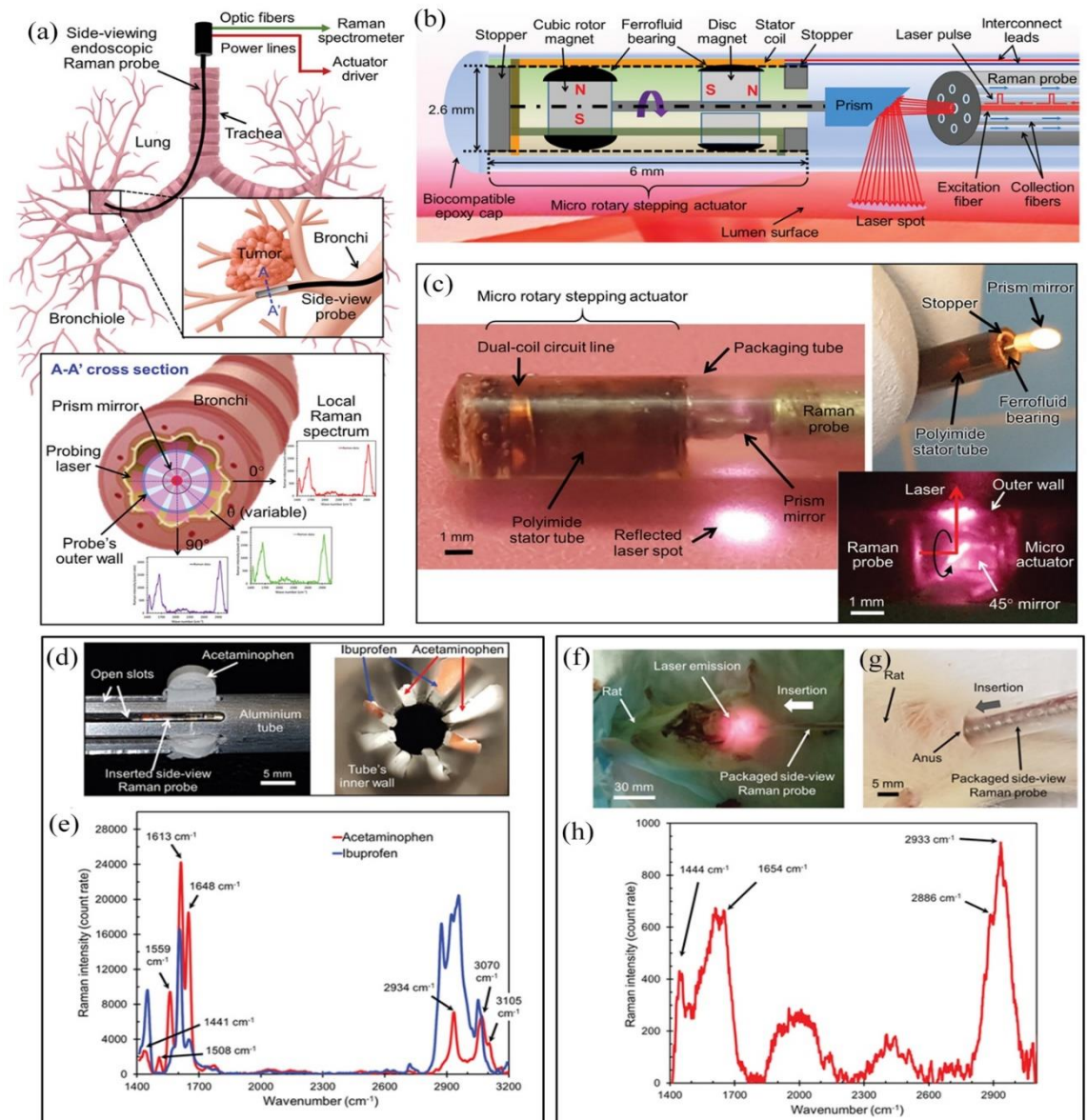

Figure 4. Schematic overview of the side-viewing endoscopic Raman spectroscopy device, angle-resolved chemical detection test and in situ test. (a) Conceptual schematics of the side-viewing Raman device developed with intended application in peripheral lung cancer detection. (b) Schematic diagram of the construction of the side-view mechanism. (c) An example of the integrated side-viewing Raman spectroscopy device (left) circumferentially scanning a probing laser beam by the prism mirror (bottom right) and the pre-assembly stator tube with the rotor-mirror component showing the ferrofluid bearing inside (top right). (d, e) Angle-resolved chemical detection test. Tubular test set-up with acetaminophen and ibuprofen samples alternately loaded in the open slots, and an axial internal view of the set-up with the two test chemicals. The Raman signals of acetaminophen and ibuprofen samples were obtained stepwise, showing high accuracy of the measured peak positions. (f, g, h) In situ test using a rat colon model. The side-view Raman probe was inserted into the colon of the euthanized rat indicating emissions of the probing laser (f), the side-viewing probe inserted to colon through anus and rectum (g) and Raman spectrum obtained from the raw data (h). Reproduced from Ref. [153] with permission from Wiley Online Library, copyright 2019.

Another popular area of research that applies Raman technology to the CRC diagnostic process is liquid biopsy. Raman spectroscopy has exceptional potential for use in the analysis of biofluids for several reasons. One is water, which is the major component of all biofluids and is a very weak Raman scatterer. Therefore, many recent studies have focused on achieving CRC detection using the Raman signals of biomarkers in biofluids. Long et al. [154] employed gold nanoparticle colloids mixed with serum from patients with CRC to provide strong and stable SERS profiles. Their spectral analysis supported that patients with CRC have lower serum concentrations of tyrosine and hypoxanthine than healthy volunteers, which is in accordance with earlier metabolomic studies [155, 156]. However, the appearance of characteristic peaks is not directly associated with cancerous lesions, and precancerous lesions intermediate between healthy tissue and CRC were not considered in this study. Thus, once samples of precancerous lesions or other hyperplastic tissues are introduced, the accuracy of the proposed classification method will be negatively affected. In another study, Feng et al. [157] utilized SERS to analyze blood plasma to detect CRC considering adenomatous polyps. After generating the classification model with the PLS-DA method, the proposed model achieved a diagnostic sensitivity of 86.4% and specificity of 80.0% for CRC and polyps, respectively, suggesting that biomolecular differences exist among the blood plasma samples of the CRC group, adenomatous polyp group, and regular volunteers. Moreover, the Raman spectroscopy of blood samples provides more information than the diagnostic basis for CRC. Both point mutations and deletions play a significant role in tumorigenesis, promotion, invasion, and metastasis of cancer, as well as in chemotherapy resistance [158]. The PCR-assisted Raman system makes performing genetic testing on tissue samples possible in order to explore individual drug resistance and guide the choice of treatment options. Li et al. [159] developed a procedure based on PCR and SERS by amplifying DNA-containing target mutations and annealing probes to detect six mutations located in *BRAF, KRAS,* and *PIK3CA* in plasma samples from 49 patients with CRC. Although only two specific mutations were related to right-sided colon cancer in this preliminary study with a small sample size, and no extended discussion on associations between genetic mutations and individuals exist, the combination of PCR and Raman is still a prospective revelation because it allows early diagnosis before tissue lesions and demonstrates the potential to perform chemotherapy drug selection.

## 3.2 Infectious diseases

Infections caused by bacterial pathogens, viruses, and parasites are commonly encountered in clinical settings and are considered the top 10 most common causes of death globally [160]. In low-income countries and regions, sepsis, a disorder characterized by systemic inflammation secondary to infection, is a substantial contributor to mortality [161]. Herein, we summarize the latest advances in Raman technology for diagnosing and evaluating infectious diseases.

### 3.2.1 Pathogenic bacteria

Overall, Raman spectroscopy provides a whole-organism fingerprint, which is typically referred to as information on the chemical compositions and biomolecular structures during bacterial sample analysis [162]. Currently, owing to the low concentration of bacteria in clinical samples, the majority of studies performing Raman spectroscopy on clinical bacterial pathogens require enrichment of disease-causing bacteria, and the most common practice is culture on agar plates [163]. However, as in other conventional detection methods for bacterial pathogens, which rely on medium culture [164, 165], this inevitably takes a long time. Moreover, only few bacteria could be successfully cultured because of the growth requirements, which may significantly affect the accuracy of the experimental results. To overcome the limitations of time wasting, attempts to apply Raman spectroscopy to tissues for in situ diagnosis of infectious diseases have increased [166]. The application of SERS is a common method to improve the signal-to-noise ratio, which enhances the intensity of the Raman signal and can be used as an alternative to concentrated culture of bacteria. Kelly et al. [167] have reported an SERS-based system for the detection of chemisorbed methyl sulfide in the headspace of six live bacterial cultures (*Escherichia coli* DH5α, *E. coli* K12 WT, *Staphylococcus aureus* Cowan I, *Enterococcus faecalis* ATCC 10541, *Pseudomonas aeruginosa* OA1, and *Bacteroides fragilis* NCTC 9343) (Figure 5a) before and after the use of antibiotics (Figure 5b–i). Since chemisorbed methyl sulfide can only be produced by living bacteria metabolizing dimethyl disulfide, the efficacy of antibiotics can be evaluated using multiple analyses. Although this study did not simulate the biochemical environment, it is still a promising step toward the bedside detection of bacterial infections and rapid antibiotic drug testing. Moreover, culture-free identification of bacterial pathogens has been developed [168] to identify antibiotic-susceptible bacteria from antibiotic-resistant bacteria and diagnose mixed flora infections. Unfortunately, although SERS is considered an excellent analytical technique, it has not yet been used as a routine diagnostic method, owing to some limitations. The problem receiving the most attention is the fabrication of suitable substrates with unique features [169]. In addition to bacterial infections, Raman spectroscopy has been applied to identify other microbial species, which has inspired the accurate diagnosis of parasites and viruses [170].

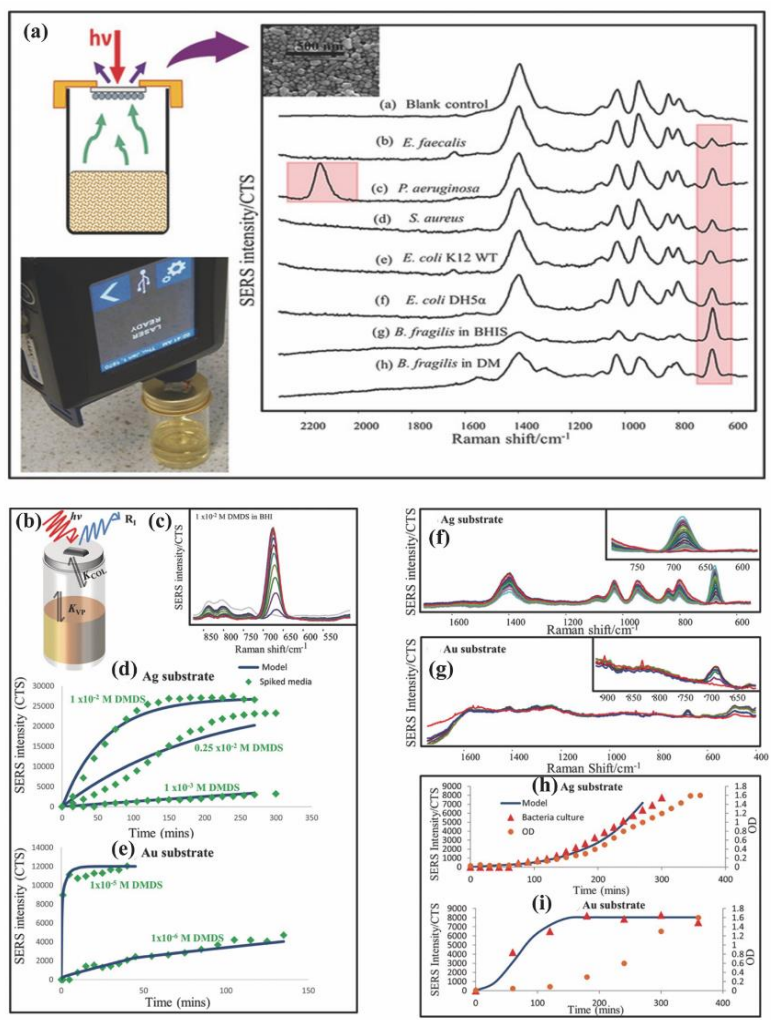

Figure 5. Inset: SEM image of the surface-enhanced Raman spectroscopy (SERS) substrate. The top image is a schematic of the detection principle and configuration for headspace SERS measurements. The photograph illustrates the experimental system used for in situ monitoring of the culture headspace at 37 °C. SERS spectra of the headspace of a) blank control, b) *E. faecalis* ATCC 10541, c) *P. aeruginosa* OA1, d) *S. aureus* Cowan I, e) E. coli K12 WT, f) E. coli DH5α, g) *B. fragilis* NCTC 9343 cultivated in brain heart infusion broth supplemented (BHIS), and h) *B. fragilis* NCTC 9343 cultivated in defined medium (DM). (b) Illustration of the relationship between the equilibration of (dissociation of dimethyl disulfide) DMDS into the headspace and the adsorption of methyl sulfide onto the enhancing substrate. (c) Headspace SERS spectra of the time-dependent adsorption from BHI broth spiked with $1\times10^{-2}$ m DMDS.24 Spectra recorded at 15 min intervals with Ag substrate. (d,e) Growth of methyl sulfide SERS signal with time for BHI broth spiked with different concentrations of DMDS using Ag substrate (d) and Au substrate (e). Solid lines fit to data from the kinetic model discussed in the text. Accumulation times were 1 s and 20 s for Ag and Au substrates, respectively. Uncertainty (S.D.) in SERS intensity was 4.8 % for Ag and 9 % for Au. (f–i) SERS spectra of kinetic run of the adsorption of methyl sulfide from *E. coli* culture on f) Ag substrate and g) Au substrate. (h, i)

Comparison of the kinetic data showing bacterial growth (OD) and increasing methyl sulfide SERS intensity with time for Ag (h) and Au (i) substrates. Solid lines fit the SERS data using the kinetic model discussed in the text. Uncertainty (S.D.) in SERS intensity was 4.8 % for Ag and 9 % for Au. Reproduced from ref. [167] with permission from Wiley Online Library, copyright 2018.

### 3.2.2 Pathogenic virus

The transmission of pathological viruses is an enduring public health issue. Considering the historical events in the public health field in the recent decades, arriving at such a conclusion is expected: virus-related disease deemed containable can quickly escalate into a global calamity, such as the severe acute respiratory syndrome coronavirus 1 (SARS-CoV-1) and COVID-19 pandemics in 2003 and 2019, respectively.

COVID-19 caused by SARS-CoV-2 has been declared a public health emergency of international concern by the World Health Organization [171]. The conventional samples commonly used for diagnostic testing of SARS-CoV-2 are nasopharyngeal and oropharyngeal swabs. However, since the sample collection process is prone to discomfort, recent studies have suggested that saliva could be an effective alternative [172]. Azzi et al. [173] analyzed saliva samples from COVID-19 patients using reverse transcription-PCR (RT-PCR). All the participants tested positive, demonstrating that saliva could be a valuable sample for COVID-19 diagnosis. Although RT-PCR has been successfully used to detect COVID-19, this method has a low sensitivity for SARS-CoV-2 detection before the onset of symptoms [174]. Therefore, developing novel detection methods is crucial.

Carlomagno et al. [175] proposed a Raman spectroscopy-based saliva analysis method to differentiate SARS-CoV-2 infections. The deep-learning-based Raman classification model applied in this study could distinguish patients with >95% accuracy. Ember et al. [107] used innovative multi-instance learning-based machine learning and droplet segmentation methods to analyze the spectral data of salivary droplets. Experiments demonstrated that Raman spectroscopy could detect biomolecular changes in COVID-positive and COVID-negative saliva supernatants. Huang et al. [176] developed a deep learning model based on residual neural networks to assist SERS analysis, which was used to detect SARS-CoV-2 antigen in pharyngeal swabs or sputum of COVID-19 patients. Figure 6(a) presents the detection process of this study, and the detailed model structure is demonstrated in Figure 6(b). The diagnostic accuracy of this method was 87.7%, which proves to be a promising approach to facilitate the rapid on-site diagnosis of SARS-CoV-2.

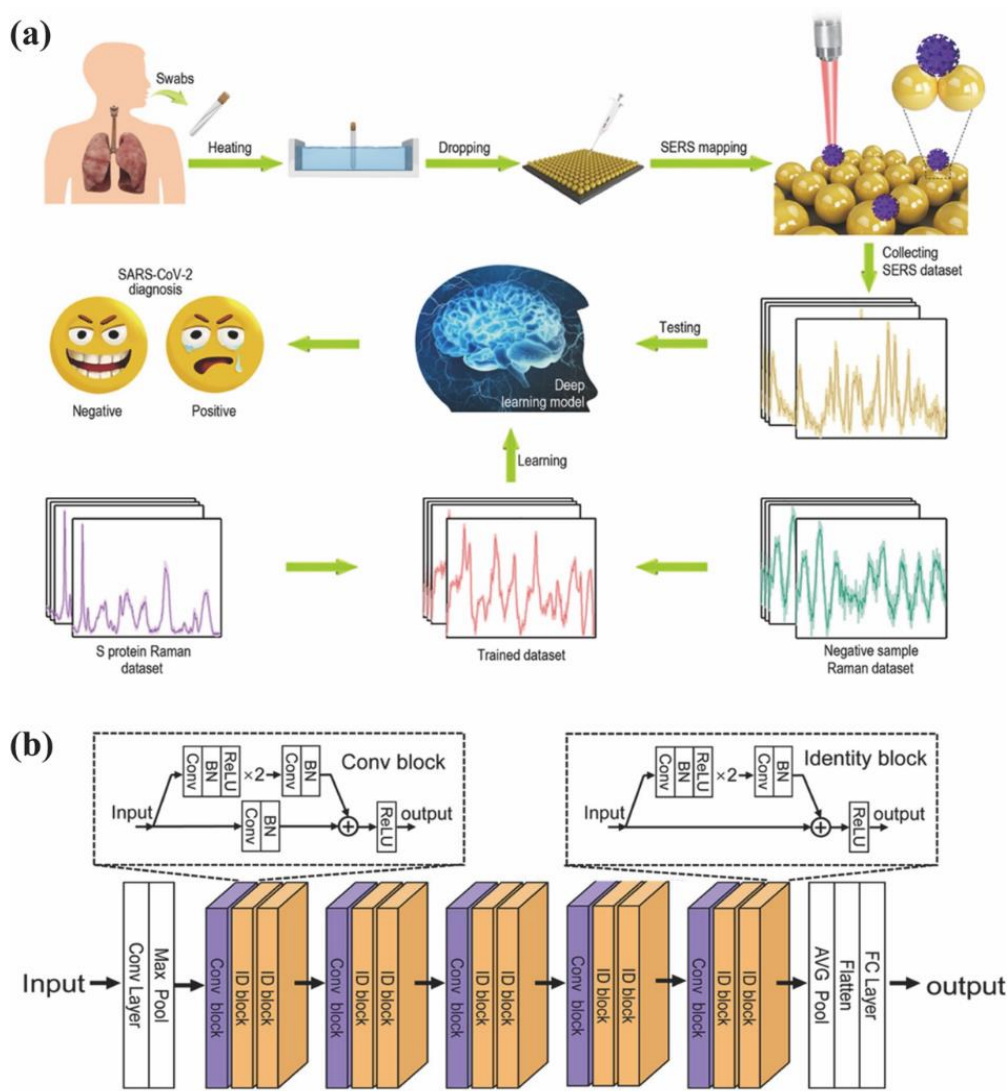

Figure 6. (a) Schematic illustration of the coronavirus disease (COVID-19) detection process of the deep learning-based surface-enhanced Raman spectroscopy (SERS) technique. (b) The architecture of the deep learning model based on residual neural networks for detecting severe acute respiratory syndrome coronavirus 2 (SARS-CoV-2). Adapted from Ref. [176] with permission from ACS Publications, copyright 2021.

Additionally, serum has been investigated as a test sample [177, 178]. Goulart et al. [179] analyzed the Raman spectra of COVID-19-positive and normal human serum samples. The discriminant analysis method used in this experiment classified the spectra with 87% sensitivity and 100% specificity. In a recent study, Paria et al. [180] presented a new platform that combined SERS and machine learning to detect SARS-CoV-2 in large areas without labels. They performed rapid detection of SARS-CoV-2 in a label-free manner by recording the SERS features of viruses on rigid and flexible substrates using plasma-active nanopatterning. This method provided test results within 25 min. Furthermore, SARS-CoV-2 continues to mutate to evade the response of the human immune system, exacerbating the challenge of controlling the COVID-19 pandemic [181]. Pezzotti et al. [182] identified significant vibrational differences between the Raman spectra of two British variant subtypes (QK002 and QHN001) present in Japan and the Japanese isolate (JPN/TY/WK-521), as illustrated in Figure 7(a–c). The authors also used customized barcodes (Figure 7(d–g)) for Raman spectroscopy to represent viral variants to aid in electronic record keeping and translate molecular features into instantly accessible information for users. This work illustrates that Raman spectroscopy could provide a clear understanding of viral structure at the molecular scale, providing researchers with timely information on SARS-CoV-2 variants and their subtypes.

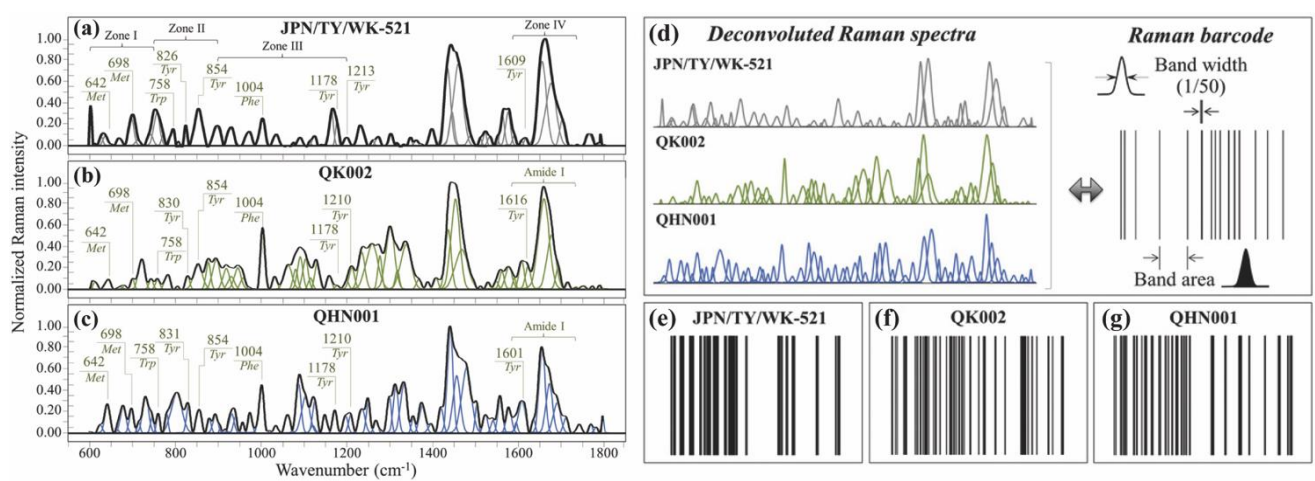

Figure 7. (a–c) Raman spectra in the frequency interval 600–1800 cm$^{-1}$ of the (a) original Japanese isolate JPN/TY/WK-521, (b) variant QK002, and (c) variant QHN001. Labels demonstrate frequencies at the maximum of selected bands (Met, Tyr, and Phe are abbreviations for methionine, tyrosine, and phenylalanine, respectively). (d) Sequences of Raman Gaussian–Lorentzian bands as deconvoluted from average Raman spectra recorded on the original Japanese isolate and two British variants (see labels), and algorithm to convert a band sequence into a barcode. (e–g) Barcodes that are constructed for the sub-band sequences in (a). Adapted from Ref. [182] with permission from Wiley Online Library, copyright 2022.

Raman spectroscopy can be used to profile nucleic acids and proteins of contaminating viruses without pretreatment. Some early studies have investigated viral RNA, DNA, and proteins from Raman spectra [170]. These studies demonstrate the extraordinary advantages of dispersive Raman techniques. The dispersive Raman system combines various excitation wavelengths and Raman microscopes, and has been applied to probe multiple pathogenic viruses. However, as an organism with a more straightforward structure than other microorganisms, most viruses share similar structures. Thus, the Raman signals collected by these systems are similar in most cases, and detecting their differences is difficult. In addition to the necessary statistical analysis, indirect viral analysis can be performed by detecting surrogates or reporter molecules instead of viruses. Typically, these techniques require physical pairing of the virus and the reporter to form a nanotag.

Adenoviruses are double-stranded DNA viruses without envelopes. They may cause acute respiratory disease, pneumonia, acute follicular conjunctivitis, cystitis, and gastroenteritis [183]. Moor et al. [184] achieved the early detection of viral infection in live human cells using Raman techniques. In this study, the viral vector Ad-CMV-control (AdC), which lacks the E1 gene coding for an early polypeptide, was used as the target virus, and human embryonic kidney 293 cells were introduced to possess the E1 gene, which contains the promoter of the E2 gene for a DNA polymerase that is specific to replication of the viral genome. The E2 peptide characteristic peaks indicated that the E2 gene was quickly transferred into the nucleus after virus invasion. This study established that the detection threshold for SERS can be reduced to a single molecule.

Similarly, this alternative diagnostic platform using SERS nanotags has been widely applied to other types of viruses. For example, the Ebola virus has been known to cause a significant epidemic in African countries in recent years. In several studies, Sebba et al. have proposed a particle-based sandwich immunoassay to detect and differentiate infections with Ebola from other more common febrile diseases (malaria and Lassa fever virus). Sebba et al. [185] created a stable optical signal by encapsulating the SERS-active Raman reporter between a 60-nm gold nanoparticle, which acts as the nanotag core, instead of attempting to detect proteins directly using SERS signals. This prospective study achieved 90.0% and 100.0% sensitivity and 97.9% and 99.6% specificity for the Ebola virus and malaria, respectively, in blood samples from non-primate animals, indicating the potential of SERS technology as an essential tool for clinical triage in low-resource settings. Recently, more evidence has demonstrated the wide range of prospects of Raman techniques for detecting viral infections, especially regarding cross-species viruses, such as the avian influenza virus [186] and Rift Valley fever virus [187, 188]. These outstanding achievements will break the limits of laboratory resources, circumvent the requirement for dedicated personnel, and make an exceptional contribution to the future of epidemiological testing.

### 3.3 Neurological diseases

Neurodegenerative diseases are a group of conditions in which neurons of the brain and spinal cord lose their function [189], and dementia is one of the most common features of these diseases. Currently, approximately 50 million people have dementia [190], which is estimated to increase to 130 million by 2050 [191]. Additionally, this group of diseases has no viable treatment, although early intervention may help delay the development process. Furthermore, previous studies have reported that relevant polymorphs accumulate in the brain tissue 10–30 years before the onset of dementia [192]. Therefore, the pre-onset detection of disease trends may be a direction for future clinical diagnosis.

### 3.3.1 Alzheimer's disease

Alzheimer's disease (AD) is the most prevalent neurodegenerative disease, affecting millions of patients worldwide, with a continuously increasing incidence [191. Although the pathogenesis of AD remains controversial, several polymers in the cerebrospinal fluid and blood have been proven to be associated with it, including neurofilament light chain, neuron-specific enolase, heart fatty acid-binding protein, chitinase-3-like protein 1, and visinin-like protein 1 [192]. Since AD is commonly detected when irreversible damage in the brain has occurred, and the definitive diagnosis can only be made post-mortem, upon the identification of the aggregates, a sensitive and affordable method for diagnosing AD represents one of the main challenges in the field.

Among many spectral analysis tools, Raman spectroscopy demonstrates high potential for identifying early-stage AD and its discrimination from late forms of AD [193, 194], as it is well suited for analyzing water-containing samples. However, the development of Raman techniques as diagnostic tools is challenging. First, the Raman signals of biomolecules are generally weak, and frequent fluorescence interference hinders the promotion of new technological breakthroughs. Several signal enhancement techniques have been used to overcome these limitations, including SERS [195, 196], TERS [197-198], CARS [199], and SRS [200, 201]. These optimized Raman-related diagnostic techniques have been commonly used in the latest prospective studies for detecting different samples.

As previously stated, blood and cerebrospinal fluid are the most frequently analyzed samples, as they may be the most biologically significant body fluids for AD. For example, Ryzhikova et al. [194] explored the potential of Raman spectroscopy in combination with machine learning to differentiate between CSF samples obtained from patients with AD and healthy controls. They explained the reasons for the characteristic bands associated with amino acid metabolism. This study achieved acceptable accuracy (84%), but its specific efficiency was further validated owing to the small sample size. Similarly, based on the classification of patients with AD and healthy individuals, Paraskevaidi et al. [193] attempted to differentiate the biochemical composition of the blood plasma of early stage AD, late-stage AD, and patients with dementia with Lewy bodies (Figure 8). The achievement of AD staging and differentiation between the two dementias may indicate that some future uses of Raman techniques could be used to detect prodromal or prognostic cases for early appropriate medical interventions.

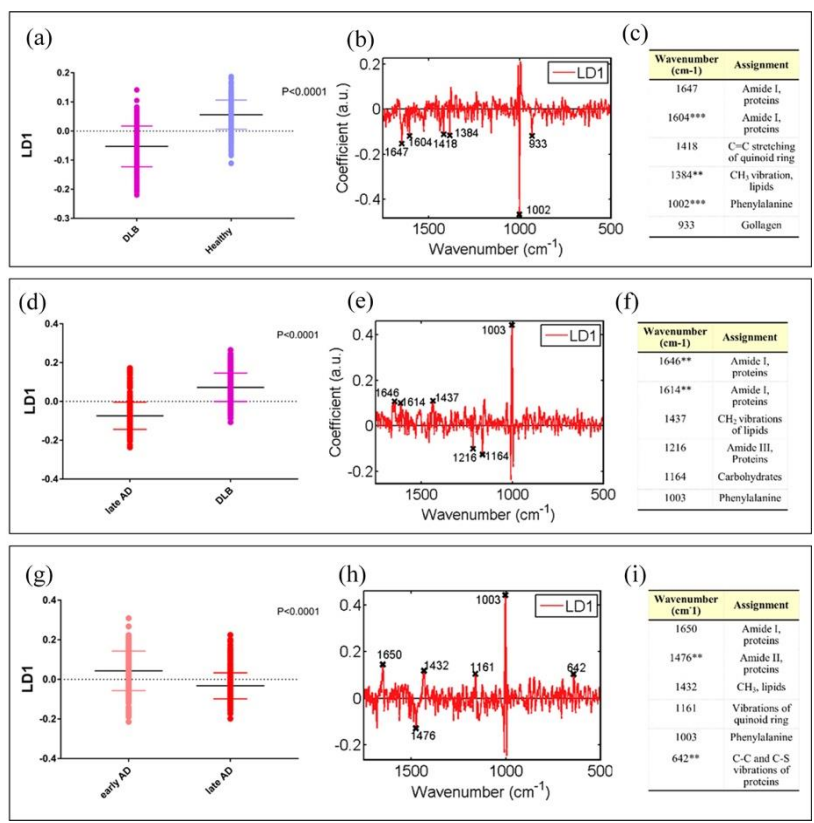

Figure 8. One-dimensional (1D) scores plot after cross-validated principal component analysis, linear discriminant analysis (PCA-LDA) (P < 0.0001, 95% CI = 0.138-0.1596) of early stage AD versus dementia with dementia with Lewy bodies (DLB) (a–c), late stage AD versus DLB (d–e), and early stage AD versus late AD, with loadings plot demonstrating the top six discriminatory peaks (left column), important peaks along with their tentative assignments (middle column), and the mean ± standard deviation (P ⩽ 0.05 was considered significant; P < 0.05 (*) or P < 0.005 (**) or P < 0.0005 (***) (right column). Reproduced from Ref. [193] with permission from ACS Publications, copyright 2018.

Several studies have also been dedicated to determining the relationship between neurotoxicity and the structures of different forms of Aβ$_{42}$ [202, 203]. Banchelli et al. [203] inspected the most superficial layers of Aβ species to identify structural motifs that are characteristic of toxic forms using SERS coupled with an intertwined silver nanowire (AgNW)-based platform. They summarized the chemical structural differences between cytotoxic Aβ$_{42}$ (A+ oligomers) and non-cytotoxic forms of Aβ$_{42}$ (A− oligomers). The authors established a connection between the exposure of Tyr and Lys residues and toxicity of Aβ$_{42}$ oligomers (Figure 9). Additionally, the secondary structures of other amyloid-β peptides were inspected by SERS to obtain information on the structural rearrangement processes involved in Raman signals, which contributes to clarifying the mechanism of AD [197, 198, 204].

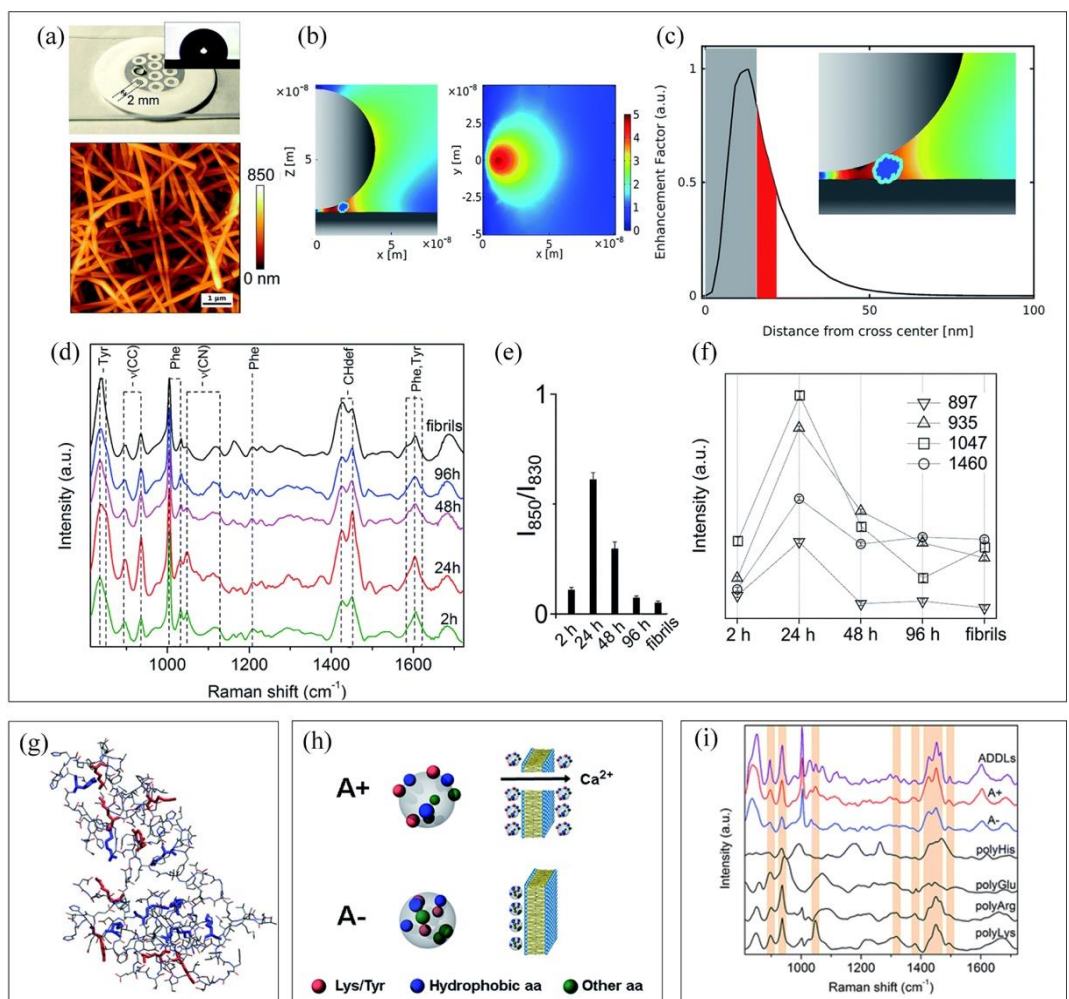

Figure 9. Surface-enhanced Raman spectroscopy (SERS) analysis of Aβ$_{42}$ species. (a) Picture of the silver-spotted substrate used for SERS analysis demonstrating a drop of Aβ$_{42}$ solution deposited on a 2-mm large spot. Inset: contact angle image of water drops after deposition on the spot; exemplary AFM image of the spot exhibiting intertwined AgNWs over the poly tetra fluoroethylene (PTFE) support. 2D sections xz in (b left) and xy in (bright) of atomic force microscope (FEM) simulations of the E-field intensity (|E|/|E0|) in-between two crossed AgNWs in air. (c) Profile of EF along the x-direction. Access to the gray zone adjacent to the origin of the intersection, from 0 to ~15 nm, is denied to the oligomer due to steric impediments, while the zone highlighted in red indicates the enhancement factor (EF) decrease experienced by a 4-nm molecule (i.e., the average dimension of an Aβ$_{42}$ oligomer) as close as possible to the hot spot (as displayed in the inset). (d) Series of SERS spectra of Aβ$_{42}$

oligomers over 2 h, 24 h, 48 h, and 96 h incubation time and of mature fibrils (from bottom to top) in the 800–1750 cm$^{-1}$ range; characteristic bands are identified with dashed lines. Analysis of selected SERS bands after multipeak fitting by Lorentzian functions: (e) ratio of 850 cm$^{-1}$ and 830 cm$^{-1}$ band intensities of Tyr doublet and (f) integrated area values of bands at 897 cm$^{-1}$, 935 cm$^{-1}$, 1047 cm$^{-1}$, and 1460 cm$^{-1}$ from spectra in (d). (g) Representative structure of compact (R < 0.9) Aβ$_{42}$ oligomer. (h) Schematic picture of the A+ and A− oligomers of Aβ$_{42}$. Toxic A+ oligomers are characterized by exposure of hydrophobic clusters (blue), as previously identified for this and related protein systems, as well as of Tyr and Lys residues (red), which is the major finding of the present study. Toxic A+ species can penetrate the cell membrane causing an influx of Ca$^{2+}$ ions from the extracellular space to the cytosol. (i) SERS spectrum of amyloid-beta derived diffusible ligands compared to that of type A+ and A− oligomers. SERS spectra of polyHis, polyGlu, polyArg and polyLys are also displayed for comparison. Bands of polyLys and/or polyArg describing relevant spectral features of type A+ oligomers and ADDLs are identified with colored boxes. Reproduced from Ref. [203] with permission from Royal Society of Chemistry, copyright 2020.

These studies demonstrate the potential of Raman techniques for enabling early AD diagnosis. However, recent studies have explored the earlier stages of self-aggregation and fibrillation [198, 204]. All of these studies constitute a complete, full-stage diagnostic system that will elucidate the pathogenesis of AD in the future and benefit patients.

### 3.3.2 Parkinson's disease

Parkinson's disease (PD) is the second most relevant neurodegenerative disorder after AD, with an increasing burden on an aging society [205]. Like AD, PD also results from the degeneration and loss of function of neurons in the brain and peripheral nervous system. It is often diagnosed through the presence of neuritic plaques and α-synuclein, which is considered the most promising biomarker. Therefore, many Raman-based studies have been conducted to explore this biomarker. Carlomagno et al. [206] investigated the relationship between PD and EVs. The latter is one of the vehicles and critical players in the transfer of α-syn throughout the body, playing a significant role in α-syn transfer across the blood-brain barrier. They characterized and analyzed by Raman spectroscopy the EV isolated from the blood of patients with PD and established a classification model to distinguish between healthy subjects and patients with PD. The reported model verified the predicted relationship, suggesting that blood-derived EVs from patients with PD have a biochemical signature that can be related to current clinical diagnosis standards. Although this finding allows the diagnosis and stratification of patients with PD, the conclusion is still limited, and the population considered is limited. To date, no evidence has confirmed that EVs are not associated with any other neurodegenerative diseases, especially those with close causes of PD. Similarly, another study has proposed a model based on Raman signals with three different types of neurodegenerative symptoms (PD, AD, and mild cognitive impairment) [207]. The differences between each group were statistically significant, and those between samples in the same group can be explained by drug treatment, inspiring both the differentiation of neurodegenerative diseases and exploration of optimal dosage of medication.

Moreover, some early studies have confirmed that α-synuclein might reduce dopamine levels in the blood of patients with PD by mediating selective deletion of dopaminergic neurons [208, 209]. Therefore, L-dopa and dopamine-receptor agonists are often used to improve dopamine deficiency in the serum for the treatment of PD [210, 211]. This advancement has greatly inspired the use of Raman technology to investigate dopamine as a biomarker for diagnosing PD, including concentration measurement [212] and exploration of adsorption mechanisms [213]. Other studies have attempted to detect and image dopamine in the presence of interfering species in animal models [213]. Ren et al. [214] have reported a SERS-based dopamine imaging approach in the cells and retinal tissues of guinea pigs and mouse models (Figure 10). The functionalized gold nanoparticle probes (AuNPs) were citrate-capped with a diameter of 40 nm, surface-modified, and stabilized with a mixture of three thiolated molecules: N-butylboronic acid-2-mercaptoethylamine, N-hydroxysuccinimide ester, and 3-sulfanylpropanenitrile, which react with the two hydroxyl groups and amine group of dopamine, respectively (Figure. 10a). Dopamine triggered the aggregation of AuNPs and the formation of plasmonic hotspots, which strongly enhanced Raman signals.

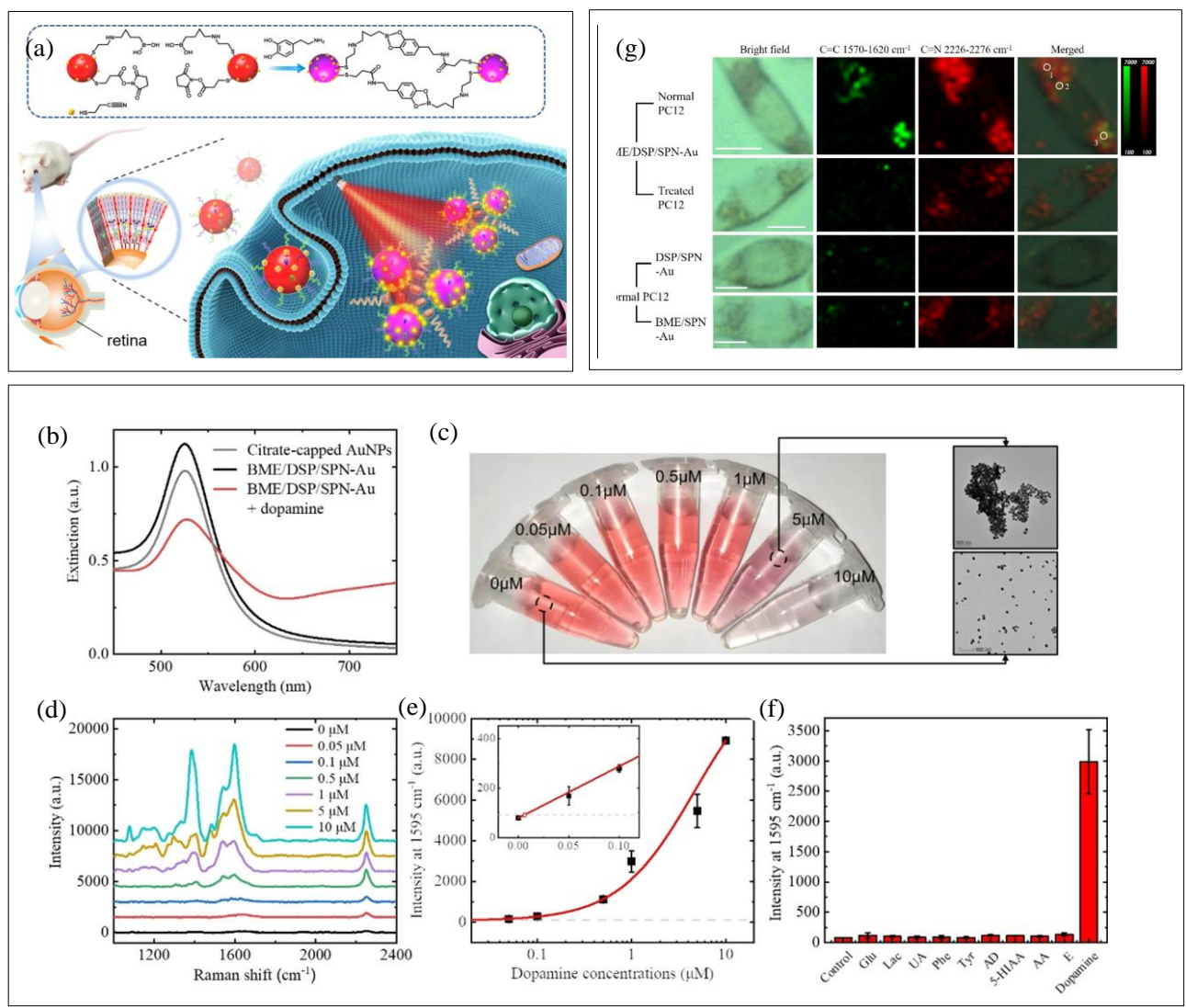

Figure 10. Schematic illustration of the surface-enhanced Raman spectroscopy (SERS) assay for local dopamine imaging based on functional AuNPs (a), which enables the formation of hotspots for subsequent cellular and retinal imaging. (b–f) Characterization of dopamine detection using BME/DSP/SPN-Au probes. (b) UV–vis spectra of the citrate-capped AuNPs (gray), the functionalized BME/DSP/SPN-Au probes before (black) and after (red) addition of dopamine (5 μM). (c) Photograph of the nanoprobes upon the addition of dopamine from 0 to 10 μM, the corresponding transmission electron microscopy (TEM) images as indicated. (d) SERS spectra of the nanoprobes upon the addition of different concentrations of dopamine (0–10 μM). (e) Plot of intensities at 1595 $cm^{-1}$ vs different concentrations of dopamine, fitted with a red curve. A broken gray line represents three times of standard deviation for LOD estimation. (f) Histogram of Raman intensities at 1595 $cm^{-1}$ in the presence of different substances at 1 μM. (g) Bright-field, SERS mapping images of differently treated single living cells acquired in 1570–1620 and 2226–2276 $cm^{-1}$ channels and their merged images. Reproduced from Ref. [214] with permission from ACS Publications, copyright 2021.

As presented in Figure 10, the detection limit was measured by adding various concentrations of dopamine (0–10 μM) to the AuNPs. The color of the functionalized AuNPs gradually changed from wine red to purple in a concentration-dependent manner because of induced particle aggregation, which was confirmed by transmission electron microscopy (Figure 10c). The UV–vis spectra demonstrated the same trend as the color changes. The SERS spectra were similar to those of the colorimetric assay (Figure 10d). This prospective study at the cellular level provides potential for further in vivo dopamine tests considering the cytotoxicity of AuNPs and laser safety.

## 4. Challenges and outlook

With the capability of label-free and highly sensitive biomolecule analysis, Raman scattering-based techniques offer robust tools for clinical diagnosis. Despite some challenges, the strength of Raman scattering will lead to the development of applications with a broader diagnostic scope, covering more lesions of different tissues. We anticipate several promising directions in the future. First, the lesion process is often accompanied by many complex biochemical reactions; however, current studies have clarified only a few reactions. In such cases, more specific molecular markers should be identified for different lesions in human tissue samples, and the relationships between markers and certain diseases need to be established. Second, as with most spectroscopic techniques, Raman spectroscopy requires advanced data processing to extract meaningful information from spectra. Machine learning has provided an unprecedented opportunity to extract information from complex or extensive spectroscopic datasets. However, the performance of machine-learning models is highly dependent on the quality of the features fed into the model for classification, which implies that machine learning requires extensive feature extraction and selection when processing large-scale data. As spectroscopy advances and spectral data sets increase, deep learning, which has achieved autonomous feature extraction, will be the focus of research on spectral processing methods. Additionally, most existing studies have been performed using known lesion samples and normal samples to make the distinction, although when Raman technology becomes a common tool for clinical diagnosis, the parameters associated with pathological diagnosis need to be standardized.

**Data availability statement**

No new data were created or analyzed in this study.

**Acknowledgments**

We acknowledge the partial support of this work by the Macau Science and Technology Development Fund (FDCT Grants 0106/2020/A3 and 0031/2021/ITP).

**Conflict of interest**

The authors declare no conflict of interest.

**CRediT author statement:**

**Yaping Qi:** Conceptualization, funding acquisition, supervision, writing–original draft, writing–reviewing, and editing. **Esther Xinyi Chen:** Writing–original draft, writing–review, and editing. **Dan Hu**: Writing–reviewing and editing. **Ying Yang**: Writing–reviewing and editing. **Zhenping Wu**: Writing–reviewing and editing. **Ming Zheng**: Writing–reviewing and editing. **Mohammad A. Sadi**: Writing–reviewing and editing. **Yucheng Jiang:** Conceptualization, supervision, reviewing, and editing. **Kang Zhang:** Conceptualization, supervision, reviewing, and editing. **Zi Chen:** Conceptualization, supervision, reviewing, and editing. **Yong P. Chen:** Conceptualization, Funding acquisition, supervision, reviewing, and editing.

**References**

[1] Paci, E. et al. Early diagnosis, not differential treatment, explains better survival in service screening. *European Journal of Cancer* **41**, 2728-2734 (2005).

[2] Moghimi-Dehkordi, B. & Safaee, A. An overview of colorectal cancer survival rates and prognosis in Asia. *World journal of gastrointestinal oncology* **4**, 71 (2012).

[3] Meyerhardt, J. A. & Mayer, R. J. Systemic therapy for colorectal cancer. *New England journal of medicine* **352**, 476-487 (2005).

[4] Favoriti, P. et al. Worldwide burden of colorectal cancer: a review. *Updates in surgery* **68**, 7-11 (2016).

[5] Ruan, S. Likelihood of survival of coronavirus disease 2019. *The Lancet Infectious Diseases* **20**, 630-631 (2020).

[6] Asai, N. et al. Early diagnosis and treatment are crucial for the survival of Pneumocystis pneumonia patients without human immunodeficiency virus infection. *Journal of Infection and*

*Chemotherapy* **18**, 898-905 (2012).

[7] DeKosky, S. T. & Marek, K. Looking backward to move forward: early detection of neurodegenerative disorders. *Science* **302**, 830-834 (2003).

[8] Li, Q. et al. Review of spectral imaging technology in biomedical engineering: achievements and challenges. *Journal of biomedical optics* **18**, 100901 (2013).

[9] Krone, N. et al. Gas chromatography/mass spectrometry (GC/MS) remains a pre-eminent discovery tool in clinical steroid investigations even in the era of fast liquid chromatography tandem mass spectrometry (LC/MS/MS). The *Journal of steroid biochemistry and molecular biology* **121**, 496-504 (2010).

[10] Kim, J. A., Wales, D. J. & Yang, G.-Z. Optical spectroscopy for in vivo medical diagnosis—a review of the state of the art and future perspectives. *Progress in Biomedical Engineering* 2, 042001 (2020).

[11] Kumar, P. Raman spectroscopy as a promising noninvasive tool in brain cancer detection. *Journal of Innovative Optical Health Sciences* 10, 1730012 (2017).

[12] Blake, N., Gaifulina, R., Griffin, L. D., Bell, I. M. & Thomas, G. M. H. Machine Learning of Raman Spectroscopy Data for Classifying Cancers: A Review of the Recent Literature. *Diagnostics* **12**, 1491 (2022).

[13] Santos, I. P. et al. Raman spectroscopy for cancer detection and cancer surgery guidance: translation to the clinics. *Analyst* **142**, 3025-3047 (2017).

[14] Mulvaney, S. P. & Keating, C. D. Raman spectroscopy. *Analytical Chemistry* **72**, 145-158 (2000).

[15] Butler, H. J. et al. Using Raman spectroscopy to characterize biological materials. *Nature protocols* **11**, 664-687 (2016).

[16] Feng, X. et al. Raman biophysical markers in skin cancer diagnosis. *Journal of Biomedical Optics* **23**, 057002 (2018).

[17] Jadhav, S. A. et al. Development of integrated microfluidic platform coupled with Surface-enhanced Raman Spectroscopy for diagnosis of COVID-19. *Medical Hypotheses* **146**, 110356 (2021).

[18] Khan, S. et al. Analysis of dengue infection based on Raman spectroscopy and support vector machine (SVM). *Biomedical optics express* **7**, 2249-2256 (2016).

[19] Yuen, C. & Liu, Q. Magnetic field enriched surface enhanced resonance Raman spectroscopy for

early malaria diagnosis. *Journal of Biomedical Optics* **17**, 017005 (2012).

[20] Morris, M. D. & Mandair, G. S. Raman assessment of bone quality. *Clinical Orthopaedics and Related Research®* **469**, 2160-2169 (2011).

[21] Guevara, E., Torres-Galván, J. C., Ramírez-Elías, M. G., Luevano-Contreras, C. & González, F. J. Use of Raman spectroscopy to screen diabetes mellitus with machine learning tools. *Biomedical Optics Express* **9**, 4998-5010 (2018).

[22] Kochan, K. et al. Pathological changes in the biochemical profile of the liver in atherosclerosis and diabetes assessed by Raman spectroscopy. *Analyst* **138**, 3885-3890 (2013).

[23] Paidi, S. K. *et al.* Label-Free Raman Spectroscopy Reveals Signatures of Radiation Resistance in the Tumor MicroenvironmentDecoding Radiation Resistance with Raman Spectroscopy. *Cancer research* **79**, 2054-2064 (2019).

[24] Lohumi, S., Kim, M. S., Qin, J. & Cho, B.-K. Raman imaging from microscopy to macroscopy: quality and safety control of biological materials. *TrAC Trends in Analytical Chemistry* **93**, 183-198 (2017).

[25] Bonifacio, A., Cervo, S. & Sergo, V. Label-free surface-enhanced Raman spectroscopy of biofluids: fundamental aspects and diagnostic applications. *Analytical and bioanalytical chemistry* **407**, 8265-8277 (2015).

[26] Smith, R., Wright, K. L. & Ashton, L. Raman spectroscopy: an evolving technique for live cell studies. *Analyst* **141**, 3590-3600 (2016).

[27] Talari, A. C. S., Movasaghi, Z., Rehman, S. & Rehman, I. U. Raman spectroscopy of biological tissues. *Applied spectroscopy reviews* **50**, 46-111 (2015).

[28] Lizio, M. G., Boitor, R. & Notingher, I. Selective-sampling Raman imaging techniques for ex vivo assessment of surgical margins in cancer surgery. *Analyst* **146**, 3799-3809 (2021).

[29] Rubina, S. & Krishna, C. M. Raman spectroscopy in cervical cancers: an update. *Journal of cancer research and therapeutics* **11**, 10 (2015).

[30] Fu, Y., Huff, T. B., Wang, H.-W., Wang, H. & Cheng, J.-X. Ex vivo and in vivo imaging of myelin fibers in mouse brain by coherent anti-Stokes Raman scattering microscopy. *Optics express* **16**, 19396-19409 (2008).

[31] Bury, D., Morais, C. L., Ashton, K. M., Dawson, T. P. & Martin, F. L. Ex vivo Raman spectrochemical analysis using a handheld probe demonstrates high predictive capability of brain

tumour status. *Biosensors* **9**, 49 (2019).

[32] Song, D. et al. Study on the biochemical mechanisms of the micro-wave ablation treatment of lung cancer by ex vivo confocal Raman microspectral imaging. *Analyst* **145**, 626-635 (2020).

[33] Brozek-Pluska, B., Kopec, M., Surmacki, J. & Abramczyk, H. Histochemical analysis of human breast tissue samples by IR and Raman spectroscopies. Protocols discussion. Infrared Physics & Technology 93, 247-254 (2018).

[34] Li, Q., Hao, C. & Xu, Z. Diagnosis of breast cancer tissues using 785 nm miniature Raman spectrometer and pattern regression. Sensors 17, 627 (2017).

[35] Ali, S. M. et al. Raman spectroscopic analysis of human skin tissue sections ex-vivo: evaluation of the effects of tissue processing and dewaxing. *Journal of Biomedical Optics* **18**, 061202 (2012).

[36] Rangaraju, L. P. et al. Classification of burn injury using Raman spectroscopy and optical coherence tomography: An ex-vivo study on porcine skin. Burns 45, 659-670 (2019).

[37] Malini, R. et al. Discrimination of normal, inflammatory, premalignant, and malignant oral tissue: a Raman spectroscopy study. *Biopolymers: Original Research on Biomolecules* **81**, 179-193 (2006).

[38] Ibrahim, O. et al. Improved protocols for pre-processing Raman spectra of formalin fixed paraffin preserved tissue sections. *Analytical Methods* **9**, 4709-4717 (2017).

[39] Meksiarun, P. et al. Comparison of multivariate analysis methods for extracting the paraffin component from the paraffin-embedded cancer tissue spectra for Raman imaging. *Scientific reports* **7**, 1-10 (2017).

[40] Ning, T. et al. Raman spectroscopy based pathological analysis and discrimination of formalin fixed paraffin embedded breast cancer tissue. *Vibrational Spectroscopy* **115**, 103260 (2021).

[41] Happillon, T. et al. Diagnosis approach of chronic lymphocytic leukemia on unstained blood smears using Raman microspectroscopy and supervised classification. *Analyst* **140**, 4465-4472 (2015).

[42] Hobro, A. J., Konishi, A., Coban, C. & Smith, N. I. Raman spectroscopic analysis of malaria disease progression via blood and plasma samples. *Analyst* **138**, 3927-3933 (2013).

[43] Hole, A. et al. Salivary Raman spectroscopy: Understanding alterations in saliva of tobacco habitués and oral cancer subjects. *Vibrational Spectroscopy* **122**, 103414 (2022).

[44] Ge, X. et al. Study on nasopharyngeal cancer tissue using surface-enhanced Raman spectroscopy.

*Optics in Health Care and Biomedical Optics VII* **10024**, 652-657 (2016).

[45] Traynor, D. et al. Raman Spectroscopy of Liquid-Based Cervical Smear Samples as a Triage to Stratify Women Who Are HPV-Positive on Screening. *Cancers* **13**, 2008 (2021).

[46] Choi, S., Moon, S. W., Shin, J.-H., Park, H.-K. & Jin, K.-H. Label-free biochemical analytic method for the early detection of adenoviral conjunctivitis using human tear biofluids. *Analytical chemistry* **86**, 11093-11099 (2014).

[47] Hu, J. et al. Raman spectroscopy analysis of the biochemical characteristics of experimental keratomycosis. *Current Eye Research* **41**, 1408-1413 (2016).

[48] Otange, B. O., Birech, Z., Okonda, J. & Rop, R. Conductive silver paste smeared glass substrates for label-free Raman spectroscopic detection of HIV-1 and HIV-1 p24 antigen in blood plasma. *Analytical and bioanalytical chemistry* **409**, 3253-3259 (2017).

[49] Birech, Z., Ondieki, A. M., Opati, R. I. & Mwangi, P. W. Low cost Raman sample substrates from conductive silver paint smear for Raman spectroscopic screening of metabolic diseases in whole blood. *Vibrational Spectroscopy* **108**, 103063 (2020).

[50] Birech, Z., Mwangi, P. W., Bukachi, F. & Mandela, K. M. Application of Raman spectroscopy in type 2 diabetes screening in blood using leucine and isoleucine amino-acids as biomarkers and in comparative anti-diabetic drugs efficacy studies. *PLoS One* **12**, e0185130 (2017).

[51] Githaiga, J. I., Angeyo, H. K., Kaduki, K. A. & Bulimo, W. D. Chemometrics-Enabled Raman Spectrometric Qualitative Determination and Assessment of Biochemical Alterations during Early Prostate Cancer Proliferation in Model Tissue. *Journal of Spectroscopy* **2020** (2020).

[52] Cui, L., Butler, H. J., Martin-Hirsch, P. L. & Martin, F. L. Aluminium foil as a potential substrate for ATR-FTIR, transflection FTIR or Raman spectrochemical analysis of biological specimens. *Analytical Methods* **8**, 481-487 (2016).

[53] Jeong, S. et al. Fluorescence-Raman dual modal endoscopic system for multiplexed molecular diagnostics. *Scientific reports* **5**, 1-9 (2015).

[54] McGregor, H. C. et al. Real‐time endoscopic Raman spectroscopy for in vivo early lung cancer detection. *Journal of biophotonics* **10**, 98-110 (2017).

[55] Shu, C., Zheng, W., Lin, K., Lim, C. & Huang, Z. Label-free follow-up surveying of post-treatment efficacy and recurrence in nasopharyngeal carcinoma patients with fiberoptic Raman endoscopy. *Analytical Chemistry* **93**, 2053-2061 (2021).

[56] Wang, J. et al. Comparative study of the endoscope-based bevelled and volume fiber-optic Raman probes for optical diagnosis of gastric dysplasia in vivo at endoscopy. *Analytical and bioanalytical chemistry* **407**, 8303-8310 (2015).

[57] Wang, J. et al. Simultaneous fingerprint and high-wavenumber fiber-optic Raman spectroscopy improves in vivo diagnosis of esophageal squamous cell carcinoma at endoscopy. *Scientific reports* **5**, 1-10 (2015).

[58] Almond, L. M. et al. Endoscopic Raman spectroscopy enables objective diagnosis of dysplasia in Barrett's esophagus. *Gastrointestinal endoscopy* **79**, 37-45 (2014).

[59] Huang, Z. et al. In vivo detection of epithelial neoplasia in the stomach using image-guided Raman endoscopy. *Biosensors and Bioelectronics* **26**, 383-389 (2010).

[60] Matthäus, C. et al. Detection and characterization of early plaque formations by Raman probe spectroscopy and optical coherence tomography: an in vivo study on a rabbit model. *Journal of Biomedical Optics* **23**, 015004 (2018).

[61] Krishna, H., Majumder, S. K., Chaturvedi, P., Sidramesh, M. & Gupta, P. K. In vivo Raman spectroscopy for detection of oral neoplasia: a pilot clinical study. *Journal of biophotonics* **7**, 690-702 (2014).

[62] Singh, S., Sahu, A., Deshmukh, A., Chaturvedi, P. & Krishna, C. M. In vivo Raman spectroscopy of oral buccal mucosa: a study on malignancy associated changes (MAC)/cancer field effects (CFE). *Analyst* **138**, 4175-4182 (2013).

[63] Jermyn, M. et al. Intraoperative brain cancer detection with Raman spectroscopy in humans. *Science translational medicine* **7**, 274ra219-274ra219 (2015).

[64] Guze, K. et al. Pilot study: Raman spectroscopy in differentiating premalignant and malignant oral lesions from normal mucosa and benign lesions in humans. *Head & neck* **37**, 511-517 (2015).

[65] Rostron, P., Gaber, S. & Gaber, D. Raman spectroscopy, review. *laser* **21**, 24 (2016).

[66] Le Ru, E. C., Meyer, M. & Etchegoin, P. G. Proof of single-molecule sensitivity in surface enhanced Raman scattering (SERS) by means of a two-analyte technique. *The journal of physical chemistry B* **110**, 1944-1948 (2006).

[67] Zhang, K. et al. Diagnosis of liver cancer based on tissue slice surface enhanced Raman spectroscopy and multivariate analysis. *Vibrational Spectroscopy* **98**, 82-87 (2018).

[68] Cialla-May, D., Zheng, X.-S., Weber, K. & Popp, J. Recent progress in surface-enhanced Raman

spectroscopy for biological and biomedical applications: from cells to clinics. *Chemical Society Reviews* **46**, 3945-3961 (2017).

[69] Nicolson, F. et al. Multiplex imaging of live breast cancer tumour models through tissue using handheld surface enhanced spatially offset resonance Raman spectroscopy (SESORRS). *Chemical Communications* **54**, 8530-8533 (2018).

[70] Liu, S., Ma, H., Zhu, J., Han, X. X. & Zhao, B. Ferrous cytochrome c-nitric oxide oxidation for quantification of protein S-nitrosylation probed by resonance Raman spectroscopy. *Sensors and Actuators B: Chemical* **308**, 127706 (2020).

[71] Bonhommeau, S., Cooney, G. S. & Huang, Y. Nanoscale chemical characterization of biomolecules using tip-enhanced Raman spectroscopy. *Chemical Society Reviews* (2022).

[72] Sonntag, M. D., Pozzi, E. A., Jiang, N., Hersam, M. C. & Van Duyne, R. P. Recent advances in tip-enhanced Raman spectroscopy. *The journal of physical chemistry letters* **5**, 3125-3130 (2014).

[73] Verma, P. Tip-enhanced Raman spectroscopy: technique and recent advances. *Chemical reviews* **117**, 6447-6466 (2017).

[74] Wang, X. et al. Tip-enhanced Raman spectroscopy for surfaces and interfaces. *Chemical Society Reviews* **46**, 4020-4041 (2017).

[75] Krafft, C., Dietzek, B., Popp, J. & Schmitt, M. Raman and coherent anti-Stokes Raman scattering microspectroscopy for biomedical applications. Journal of biomedical optics 17, 040801 (2012).

[76] Aljakouch, K. et al. Fast and noninvasive diagnosis of cervical Cancer by coherent anti-stokes Raman scattering. Analytical chemistry 91, 13900-13906 (2019).

[77] Zhang, J. et al. in Advanced Chemical Microscopy for Life Science and Translational Medicine 2022.   29-33 (SPIE).

[78] Zhang, B. et al. Highly specific and label-free histological identification of microcrystals in fresh human gout tissues with stimulated Raman scattering. Theranostics 11, 3074 (2021).

[79] Figueroa, B. et al. Detecting and quantifying microscale chemical reactions in pharmaceutical tablets by stimulated Raman scattering microscopy. Analytical Chemistry 91, 6894-6901 (2019).

[80] Robert B. Resonance Raman spectroscopy. Photosynthesis Research 101, 147-155 (2009).

[81] Albrecht, A. C. On the theory of Raman intensities. Journal of Chemical Physics 34, 1476-1484 (1961).

[82] Kolhatkar, G., Plathier, J. & Ruediger, A. Nanoscale investigation of materials, chemical

reactions, and biological systems by tip enhanced Raman spectroscopy. Journal of Materials Chemistry C 6, 1307-1319 (2018).

[83] Kumar, N., Weckhuysen, B. M., Wain, A. J., Pollard, A. J. Nanoscale chemical imaging using tip-enhanced Raman spectroscopy. Nature protocols 14, 1169-1193 (2019).

[84] Sitjar, J. et al. Challenges of SERS technology as a non-nucleic acid or antigen detection method for SARS-CoV-2 and its variants. Biosensors and Bioelectronics 181, 113153 (2021).

[85] Jones, R. R., Hooper, D. C., Zhang, L., Wolverson, D., Valev, V. K. Raman techniques: fundamentals and frontiers. Nanoscale Research Letters 14, 1-34 (2019).

[86] van Heel, A. C. New method for transporting optical images without aberrations. *Nature* **173**, 39-39 (1954).

[87] Brenner, H. et al. Long-lasting reduction in the risk of colorectal cancer following screening endoscopy. *British journal of cancer* **85**, 972-976 (2001).

[88] Areia, M. et al. External validation of the classification for methylene blue magnification chromoendoscopy in premalignant gastric lesions. *Gastrointestinal Endoscopy* **67**, 1011-1018 (2008).

[89] Huang, Z. et al. Early in vivo diagnosis of gastric dysplasia using narrow-band image-guided Raman endoscopy. *Journal of biomedical optics* **15**, 037017 (2010).

[90] Bergholt, M. S. et al. Fiberoptic confocal Raman spectroscopy for real-time in vivo diagnosis of dysplasia in Barrett's *Gastroenterology* **146**, 27-32 (2014).

[91] Bergholt, M. et al. In vivo diagnosis of esophageal cancer using image-guided Raman endoscopy and biomolecular modeling. *Technology in cancer research & treatment* **10**, 103-112 (2011).

[92] Lin, D. et al. Autofluorescence and white-light imaging-guided endoscopic Raman and diffuse reflectance spectroscopy for in vivo nasopharyngeal cancer detection. *Journal of Biophotonics* **11**, e201700251 (2018).

[93] Kim, Y.-i. et al. Simultaneous detection of EGFR and VEGF in colorectal cancer using fluorescence Raman endoscopy. *Scientific Reports* **7**, 1-11 (2017).

[94] Desroches, J. et al. Characterization of the Raman spectroscopy probe system for intraoperative brain tissue classification. *Biomedical optics express* **6**, 2380-2397 (2015).

[95] Faulds, K., Barbagallo, R. P., Keer, J. T., Smith, W. E. & Graham, D. SERRS as a more sensitive technique for the detection of labelled oligonucleotides compared to fluorescence. *Analyst* **129**,

567-568 (2004).

[96] Lee, S. et al. Biological imaging of HEK293 cells expressing PLCγ1 by surface-enhanced Raman microscopy. *Analytical chemistry* **79**, 916-922 (2007).

[97] Huang, R. et al. High-precision imaging of microscopic spread of glioblastoma with a targeted ultrasensitive SERRS molecular imaging probe. *Theranostics* **6**, 1075 (2016).

[98] Dinish, U., Balasundaram, G., Chang, Y.-T. & Olivo, M. Actively targeted the in vivo multiplex detection of intrinsic cancer biomarkers using biocompatible SERS nanotags. *Scientific reports* **4**, 1-7 (2014).

[99] Karabeber, H. et al. Guiding brain tumor resection using surface-enhanced Raman scattering nanoparticles and a handheld Raman scanner. ACS nano 8, 9755-9766 (2014).

[100] Pence, I., Mahadevan-Jansen, A. Clinical instrumentation and applications of Raman spectroscopy. *Chemical Society Reviews* **45**, 1958-1979 (2016).

[101] Fan, Y. et al. Rapid noninvasive screening of cerebral ischemia and cerebral infarction based on tear Raman spectroscopy combined with multiple machine learning algorithms. Lasers Med. Sci. 37, 417-424 (2022).

[102] Sciortino, T. et al. Raman spectroscopy and machine learning for IDH genotyping of unprocessed glioma biopsies. Cancers. 13, 4196 (2021).

[103] Riva, M. et al. Glioma biopsy classification using Raman spectroscopy and machine-learning models of fresh tissue samples. Cancers. 13, 1073 (2021).

[104] Ralbovsky, N. M., Halámková, L., Wall, K., Anderson-Hanley, C., Lednev, I. K. Screening for Alzheimer's disease using saliva: A new approach based on machine learning and Raman hyperspectroscopy. J. Alzheimer's Dis. 71, 1351-1359 (2019).

[105] Wang, Z. et al. Rapid biomarker screening for Alzheimer's disease using interpretable machine learning and graphene-assisted Raman spectroscopy. ACS Nano. 16, 6426-6436 (2022).

[106] Yin, G. et al. Efficient primary screening of COVID-19 using serum Raman spectroscopy. J. Raman Spectrosc. 52, 949-958 (2021).

[107] Ember, K. et al. Saliva-based detection of COVID-19 infection in a real-world setting using reagent-free Raman spectroscopy and machine learning. J. Biomed. Opt. 27, 025002 (2022).

[108] Zhang, L. et al. Raman spectroscopy and machine learning for the classification of breast cancers. Spectrochim. Acta A Mol. Biomol. Spectrosc. 264, 120300 (2022).

[109]    Fallahzadeh, O., Dehghani-Bidgoli, Z., Assarian, M. Raman spectral feature selection using ant colony optimization for breast cancer diagnosis. Lasers in Medical Science 33, 1799-1806 (2018).

[110]    Ma, D. et al. Classification of breast cancer tissue using Raman spectroscopy with a one-dimensional convolutional neural network. Spectrochimica Acta Part A: Molecular and Biomolecular Spectroscopy 256, 119732 (2021).

[111]    Qi, Y. et al. Accurate diagnosis of lung tissues for 2D Raman spectrograms by deep learning based on a short-time Fourier transform. Anal. Chim. Acta. 1179, 338821 (2021).

[112]    Shin, H. et al. Early stage lung cancer diagnosis by deep learning-based spectroscopic analysis of circulating exosomes. ACS Nano. 14, 5435-5444 (2020).

[113]    Chen, F. et al. Screening ovarian cancer using Raman spectroscopy of blood plasma coupled with machine learning data processing. Spectrochim. Acta A Mol. Biomol. Spectrosc. 265, 120355 (2022).

[114]    Li, M. et al. A Novel and Rapid Serum Detection Technology for Noninvasive Screening of Gastric Cancer Based on Raman Spectroscopy Combined With Different Machine Learning Methods. Front. Oncol. 11, 665176 (2021).

[115]    He, C., Wu, X., Zhou, J., Chen, Y., Ye, J. Raman optical identification of renal cell carcinoma via machine learning. Spectrochim. Acta A Mol. Biomol. Spectrosc. 252, 119520 (2021).

[116]    Lussier, F., Thibault, V., Charron, B., Wallace, G. Q., Masson, J. F. (2020). Deep learning and artificial intelligence methods for Raman and surface-enhanced Raman scattering. TrAC Trends Anal Chem 124, 115796.

[117]    Liu, J., Osadchy, M., Ashton, L., Foster, M., Solomon, C. J., & Gibson, S. J. (2017). Deep convolutional neural networks for Raman spectrum recognition: A unified solution. Analyst, 142(21), 4067-4074.

[118]    Hulvat M.. Cancer incidence and trends. *Surgical Clinics* **100**, 469-481 (2020).

[119]    Depciuch, J. et al. Raman and FTIR spectroscopy were used to determine the chemical changes in healthy brain and glioblastoma tumor tissues. *Spectrochimica Acta Part A: Molecular and Biomolecular Spectroscopy* **225**, 117526 (2020).

[120]    Bovenkamp, D. et al. Combination of high-resolution optical coherence tomography and Raman spectroscopy for improved staging and grading of bladder cancer. *Applied Sciences* **8**, 2371 (2018).

[121]   Barroso, E. M. et al. Water concentration analysis using Raman spectroscopy was used to determine the location of the tumor border during oral cancer surgery. *Cancer research* **76**, 5945-5953 (2016).

[122]   Sung, H. et al. Global Cancer Statistics 2020: GLOBOCAN estimates of incidence and mortality worldwide for 36 cancers in 185 countries. *CA: a cancer journal for clinicians* **71**, 209-249 (2021).

[123]   Abramczyk, H. & Brozek-Pluska, B. Raman Imaging in Biochemical and Biomedical Applications. Diagnosis and treatment of breast cancer. Chemical Reviews 113, 5766-5781 (2013).

[124]   Lyng, F. M. et al. Discrimination of breast cancer from benign tumors using Raman spectroscopy. *PLoS One* **14**, e0212376 (2019).

[125]   Yala, A., Lehman, C., Schuster, T., Portnoi, T., Barzilay, R. Deep-learning mammography-based model for improved breast cancer risk prediction. *Radiology* **292**, 60-66 (2019).

[126]   In addition, Pichardo-Molina et al. Raman spectroscopy and multivariate analysis of serum samples from patients with breast cancer *Lasers in medical science* **22**, 229-236 (2007).

[127]   Lin, T., Song, Y., Liao, J., Liu, F., Zeng, T.-T. Applications of surface-enhanced Raman spectroscopy in detection *Nanomedicine* **15**, 2971-2989 (2020).

[128]   Nargis, H. et al. Raman spectroscopy of blood plasma samples from breast cancer patients at different stages. *Spectrochimica Acta Part A: Molecular and Biomolecular Spectroscopy* **222**, 117210 (2019).

[129]   Barta J. A., Powell C. A., Wisnivesky J.. Global epidemiology of lung cancer. *Annals of global health* **85** (2019).

[130]   Kong, K., Kendall, C., Stone, N., Notingher, I. Raman spectroscopy for medical diagnostics—from in vitro biofluid assays to in vivo cancer detection. *Advanced drug delivery reviews* **89**, 121-134 (2015).

[131]   Li, X., Yang, T., & Lin, J. Spectral analysis of human saliva for detection of lung cancer using surface-enhanced Raman spectroscopy. *Journal of biomedical optics* **17**, 037003 (2012).

[132]   Qian, K., Wang, Y., Hua, L., Chen, A. & Zhang, Y. New method of lung cancer detection by saliva test using surface-enhanced Raman spectroscopy. *Thoracic cancer* **9**, 1556-1561 (2018).

[133]   Ke, Z.-Y. et al. Efficacy of Raman spectroscopy in lung cancer diagnosis: first diagnostic meta-analysis. *Lasers in Medical Science* **37**, 425-434 (2022).

[134]   Zhang, C. et al. Urine proteome profiling predicts lung cancer in control cases and other tumors. *EBioMedicine* **30**, 120-128 (2018).

[135]   Mathé, E. A. et al. Noninvasive urinary metabolomic profiling identifies diagnostic and prognostic markers in lung cancer. *Cancer research* **74**, 3259-3270 (2014).

[136]   Carrola, J. et al. Metabolic signatures of lung cancer in biofluids: NMR-based metabolomics of urine. *Journal of proteome research* **10**, 221-230 (2011).

[137]   Yang, T. et al. Facile and label-free detection of lung cancer biomarkers in urine using magnetically assisted surface-enhanced Raman scattering. *Acs Applied Materials & Interfaces* **6**, 20985-20993 (2014).

[138]   Bratchenko, L. A. et al. Comparative study of multivariate analysis methods for blood Raman spectra classification. *Journal of Raman Spectroscopy* **51**, 279-292 (2020).

[139]   Guo, T. et al. Highly selective detection of EGFR mutation genes in lung cancer based on surface-enhanced Raman spectroscopy and asymmetric PCR. *Journal of Pharmaceutical and Biomedical Analysis* **190**, 113522 (2020).

[140]   Tahir, M. A., Dina, N. E., Cheng, H., Valev, V. K., & Zhang, L. Surface-enhanced Raman spectroscopy for bioanalysis and diagnosis. *Nanoscale* **13**, 11593-11634 (2021).

[141]   Liu, K. et al. Label-free surface-enhanced Raman spectroscopy of serum was based on multivariate statistical analysis for the diagnosis and staging of lung adenocarcinoma. *Vibrational Spectroscopy* **100**, 177-184 (2019).

[142]   Bukva, M. et al. Raman spectral signatures of serum-derived EV-enriched isolates may support the diagnosis of CNS tumors. *Cancers* **13**, 1407 (2021).

[143]   Wang, H. et al. Screening and staging for non-small cell lung cancer by serum laser Raman spectroscopy. *Spectrochimica Acta Part A: Molecular and Biomolecular Spectroscopy* **201**, 34-38 (2018).

[144]   Yamamoto, S. et al. Analysis of an ADTKD family with a novel frameshift mutation in MUC1 revealed the characteristic features of the mutant MUC1 protein. *Nephrology Dialysis Transplantation* **32**, 2010-2017 (2017).

[145]   Han, Q. et al. Vps4A mediates the localization and exosome release of β-catenin to inhibit epithelial-mesenchymal transition in hepatocellular carcinoma. *Cancer letters* **457**, 47-59 (2019).

[146]   Park, J. et al. Exosome classification by pattern analysis of surface-enhanced Raman

spectroscopy data for lung cancer diagnosis. *Analytical chemistry* **89**, 6695-6701 (2017).

[147] Avella-Oliver, M., Puchades, R., Wachsmann-Hogiu, S., Maquieira, A. Label-free SERS analysis of proteins and exosomes with large-scale substrates from recordable compact disks. *Sensors and Actuators B: Chemical* **252**, 657-662 (2017).

[148] In addition, Sivashanmugan et al. Bimetallic nanoplasmonic gap-mode SERS substrate for normal lung and cancer-derived exosome detection. *Journal of the Taiwan Institute of Chemical Engineers* **80**, 149-155 (2017).

[149] Shin H., Jeong H., Park J., Hong S, Choi Y, Correlation between cancerous exosomes and protein markers based on surface-enhanced Raman spectroscopy (SERS) and principal component analysis (PCA). *ACS sensors* **3**, 2637-2643 (2018).

[150] Wu, W. et al. Surface plasmon resonance imaging-based biosensor for multiplex and ultrasensitive detection of NSCLC-associated exosomal miRNAs using a DNA-programmed heterostructure of Au-on-Ag. *Biosensors and Bioelectronics* **175**, 112835 (2021).

[151] Rojalin, T., Phong, B., Koster, H. J., Carney, R. P. Nanoplasmonic approaches for sensitive detection and molecular characterization of extracellular vesicles. *Frontiers in chemistry* **7**, 279 (2019).

[152] Short M. A., Wang W., Tai I. T., Zeng H. Development and in vivo testing of a high-frequency endoscopic Raman spectroscopy system for potential applications in the detection of early colonic neoplasia. *Journal of Biophotonics* **9**, 44-48 (2016).

[153] Jayhooni, S. M. H. et al. Side‐viewing endoscopic Raman spectroscopy for angle-resolved analysis of luminal organs. *Advanced Materials Technologies* 4, 1900364 (2019).

[154] Hong, Y. et al. Label-free diagnosis of colorectal cancer through coffee ring‐assisted surface-enhanced Raman spectroscopy on blood serum. *Journal of Biophotonics* **13**, e201960176 (2020).

[155] Li, J. et al. Tyrosine and glutamine-leucine are metabolic markers of early stage cancers. *Gastroenterology* **157**, 257-259. e255 (2019).

[156] Long, Y. et al. Global and targeted serum metabolic profiling of colorectal cancer progression. *Cancer* **123**, 4066-4074 (2017).

[157] Feng, S. et al. Label-free surface-enhanced Raman spectroscopy for detection of colorectal cancer and precursor lesions using blood plasma. *Biomedical optics express* **6**, 3494-3502 (2015).

[158] Minamoto, T. & Ronai, Z. e. Gene Mutation as a Target for Early Detection in Cancer

Diagnosis. *Critical reviews in oncology/hematology* **40**, 195-213 (2001).

[159]   Li, X. et al. Surface-enhanced Raman spectroscopy (SERS) for multiplex detection of Braf, Kras, and Pik3ca mutations in the plasma of colorectal cancer patients. *Theranostics* **8**, 1678 (2018).

[160]   Bearman G. M., Wenzel R.. Bacteremias: a leading cause of death. *Archives of medical research* **36**, 646-659 (2005).

[161]   Shrestha, G. S. et al. International Surviving Sepsis Campaign Guidelines 2016: The perspective from low- and middle-income countries. *The Lancet Infectious Diseases* **17**, 893-895 (2017).

[162]   Ashton, L., Lau, K., Winder, C. L., Goodacre, R. Raman spectroscopy: Lighting up the future of microbial identification. *Future microbiology* **6**, 991-997 (2011).

[163]   For example, Kaewseekhao et al. Diagnosis of active and latent tuberculosis infection based on Raman spectroscopy and surface-enhanced Raman spectroscopy. *Tuberculosis* **121**, 101916 (2020).

[164]   Váradi, L. et al. Methods for the detection and identification of pathogenic bacteria: past, present, and future. *Chemical Society Reviews* **46**, 4818-4832 (2017).

[165]   Verroken, A. et al. Reducing time to identify positive blood cultures with MALDI-TOF MS analysis after a 5-h subculture. *European journal of clinical microbiology & infectious diseases* **34**, 405-413 (2015).

[166]   Dhankhar, D. et al. Resonance Raman spectra for in situ identification of bacterial strains and their inactivation mechanisms. Applied Spectroscopy 75, 1146-1154 (2021).

[167]   Kelly, J., Patrick, R., Patrick, S., Bell, S. E. Surface-enhanced Raman spectroscopy for the detection of a metabolic product in the headspace above live bacterial cultures. Angewandte Chemie International Edition 57, 15686-15690 (2018).

[168]   Tien, N. et al. Diagnosis of bacterial pathogens in the urine of patients with urinary tract infection using surface-enhanced Raman spectroscopy. *Molecules* **23**, 3374 (2018).

[169]   Ouyang, L., Ren, W., Zhu, L., Irudayaraj, J. Prosperity to challenges: Recent approaches in SERS substrate fabrication. *Reviews in Analytical Chemistry* **36** (2017).

[170]   Thomas Jr., G. J. Raman spectroscopy and virus research. *Applied Spectroscopy* **30**, 483-494 (1976).

[171]  Cucinotta, D. and Vanelli, M. The WHO declared COVID-19 as a pandemic. Acta Biomed. 91, 157-160 (2020).

[172]  Xu, R. et al. Saliva: Potential diagnostic value and transmission of 2019-nCoV. Int. J. Oral Sci. 12, 11 (2020).

[173]  Azzi, L. et al. Saliva is a reliable tool for detecting SARS-CoV-2. J. Infect. 81, e45-e50 (2020).

[174]  L. Kucirka et al., "Variation in false-negative rate of reverse transcriptase polymerase chain reaction-based SARS-CoV-2 tests by time since exposure," Ann. Intern. Med., 173 (4), 262 -267 (2020).

[175]  Carlomagno, C., Bertazioli, D., Gualerzi, A., Picciolini, S., Banfi, P. I., Lax, A., ... & Bedoni, M. (2021). COVID-19 salivary Raman fingerprint: An innovative approach for the detection of current and past SARS-CoV-2 infections. Scientific Reports, 11(1), 1-13.

[176]  Huang, J. et al. On-site detection of the SARS-CoV-2 antigen by deep learning-based surface-enhanced Raman spectroscopy and its biochemical foundations. *Analytical Chemistry* **93**, 9174-9182 (2021).

[177]  Han, H. et al. Profiling of serum cytokines in COVID-19 patients revealed that IL-6 and IL-10 are disease severity predictors. *Emerging microbes & infections* **9**, 1123-1130 (2020).

[178]  Rabbani G., Ahn S.. Roles of human serum albumin in prediction, diagnosis, and treatment of COVID-19. International Journal of Biological Macromolecules 193, 948-955 (2021).

[179]  Goulart AC, et al. Diagnosis of COVID-19 in human serum using Raman spectroscopy. *Lasers in Medical Science*, 1-10 (2022).

[180]  Paria, D. et al. Label-free spectroscopic SARS-CoV-2 detection on versatile nanoimprinted substrates. *Nano letters* **22**, 3620-3627 (2022).

[181]  Garcia-Beltran et al. Multiple SARS-CoV-2 variants escape neutralization by vaccine-induced humoral immunity. Cell 184, 2372-2383. e2379 (2021).

[182]  Pezzotti, G. et al. Raman Molecular Fingerprints of SARS‐CoV‐2 British Variant and the Concept of Raman Barcode. Advanced Science 9, 2103287 (2022).

[183]  Doerfler, W. in *Medical Microbiology. 4th edition*. (1996).

[184]  Moor, K. et al. Early detection of virus infection in live human cells using Raman spectroscopy. *Journal of Biomedical Optics* **23**, 097001 (2018).

[185]  Sebba, D. et al. Point-of-care diagnosis for differentiating Ebola from endemic febrile diseases.

*Science translational medicine* **10**, eaat0944 (2018).

[186]   Neng, J., Li, Y., Driscoll, A. J., Wilson, W. C. & Johnson, P. A. Detection of multiple pathogens in serum using silica-encapsulated nanotags in a surface-enhanced Raman scattering-based immunoassay. *Journal of agricultural and food chemistry* **66**, 5707-5712 (2018).

[187]   Shiratori, I. et al. Selection of DNA aptamers that bind to influenza A viruses with high affinity and broad subtype specificity. *Biochemical and Biophysical Research Communications* **443**, 37-41 (2014).

[188]   Xiao, M. et al. Ultrasensitive detection of avian influenza A virus (H7N9) using surface-enhanced Raman scattering-based lateral flow immunoassay strips. *Analytica Chimica Acta* **1053**, 139-147 (2019).

[189]   Dugger, B. N., Dickson, D. W. Pathology of neurodegenerative diseases. *Cold Spring Harbor perspectives in biology* **9**, a028035 (2017).

[190]   Nichols, E. et al. Global, regional, and national burden of Alzheimer's disease and other dementias, 1990-2016: a systematic analysis for the Global Burden of Disease Study 2016. *The Lancet Neurology* **18**, 88-106 (2019).

[191]   Prince, M. J. et al. World Alzheimer Report 2015-The Global Impact of Dementia: An Analysis of Prevalence, Incidence, Cost, and Trends.   (2015).

[192]   Gordon, B. A. et al. Spatial patterns of neuroimaging biomarker changes in individuals from families with autosomal dominant Alzheimer's disease: A longitudinal study. *The Lancet Neurology* **17**, 241-250 (2018).

[193]   In addition, Paraskevaidi et al. Raman spectroscopy was used to diagnose Alzheimer's disease and dementia with Lewy bodies in the blood. *ACS chemical neuroscience* **9**, 2786-2794 (2018).

[194]   Ryzhikova, E. et al. Raman spectroscopy and machine learning for biomedical applications: Alzheimer's disease diagnosis based on cerebrospinal fluid analysis *Spectrochimica Acta Part A: Molecular and Biomolecular Spectroscopy* **248**, 119188 (2021).

[195]   Carlomagno, C. et al. SERS‐based biosensor for Alzheimer's disease evaluation through fast analysis of human serum. *Journal of biophotonics* **13**, e201960033 (2020).

[196]   Cennamo, G. et al. Surface-enhanced Raman spectroscopy of tears: a diagnostic tool for neurodegenerative disease identification. *Journal of biomedical optics* **25**, 087002 (2020).

[197]   Tabatabaei, M., Caetano, F. A., Pashee, F., Ferguson, S. S. & Lagugné-Labarthet, F. Tip-

enhanced Raman spectroscopy of amyloid β at neuronal spines. *Analyst* **142**, 4415-4421 (2017).

[198]   Zikic, B., Bremner, A., Talaga, D., Lecomte, S., Bonhommeau, S. Tip-enhanced Raman spectroscopy of Aβ (1-42) fibrils. *Chemical Physics Letters* **768**, 138400 (2021).

[199]   Talaga, D. et al. PIP2 Phospholipid-Induced Aggregation of Tau Filaments Probed by Tip‐Enhanced Raman Spectroscopy. *Angewandte Chemie* **130**, 15964-15968 (2018).

[200]   Cunha, R. et al. Nonlinear and vibrational microscopy for label-free characterization of amyloid-β plaques in an AD model. *Analyst* **146**, 2945-2954 (2021).

[201]   Ji, M. et al. Label-free imaging of amyloid plaques in Alzheimer's disease using stimulated Raman scattering microscopy. *Science advances* **4**, eaat7715 (2018).

[202]   Banchelli, M. et al. Spot‐on SERS detection of biomolecules using laser-patterned dot arrays of assembled silver nanowires. ChemNanoMat 5, 1036-1043 (2019).

[203]   Banchelli, M. et al. Nanoscopic insights into the surface conformation of neurotoxic Aβ oligomers. *RSC advances* **10**, 21907-21913 (2020).

[204]   Garcia-Leis, A. & Sanchez-Cortes, S. Label-Free Detection and Self-Aggregation of Amyloid β-peptides Based on Plasmonic Effects Induced by Ag Nanoparticles: Implications in Alzheimer's Disease Diagnosis. *ACS Applied Nano Materials* 4, 3565-3575 (2021).

[205]   Berg D. Biomarkers for the early detection of Parkinson's and Alzheimer's disease. *Neurodegenerative Diseases* **5**, 133-136 (2008).

[206]   Carlomagno, C. et al. Identification of the Raman Salivary Fingerprint of Parkinson's Disease Using a Spectroscopic-Computational Combinatory Approach. *Frontiers in neuroscience* **15** (2021).

[207]   Ryzhikova, E. et al. Raman spectroscopy of blood serum for Alzheimer's disease diagnostics: Specificity relative to other types of dementia. *Journal of biophotonics* **8**, 584-596 (2015).

[208]   Vila, M. et al. α‐Synuclein upregulation in substantia nigra dopaminergic neurons following administration of the parkinsonian toxin MPTP. *Journal of neurochemistry* **74**, 721-729 (2000).

[209]   Masliah, E. et al. Dopaminergic loss and inclusion body formation in α-synuclein mice: Implications for neurodegenerative disorders. *Science* **287**, 1265-1269 (2000).

[210]   Dunn, A. R. et al. Synaptic vesicle glycoprotein 2C (SV2C) modulates dopamine release and is disrupted in Parkinson disease. *Proceedings of the National Academy of Sciences* **114**, E2253-

E2262 (2017).

[211]   Olanow, C. W., Schapira, A. H., Rascol, O. Continuous dopamine receptor stimulation in early Parkinson's disease. *Trends in neurosciences* **23**, S117-S126 (2000).

[212]   Cutler, E. G. *Dopamine Detection via Surface Enhanced Raman Scattering for Parkinson's Disease*, The University of North Carolina at Charlotte, (2019).

[213]   Figueiredo, M. L. et al. Surface-enhanced Raman scattering for dopamine in Ag colloid: Adsorption mechanism and detection in the presence of interfering species. *Applied Surface Science* **522**, 146466 (2020).

[214]   Ren, X. et al. Dopamine imaging in living cells and retina using surface-enhanced Raman scattering based on functionalized AuNPs *Analytical Chemistry* **93**, 10841-10849 (2021).